\documentclass[final,5p,times,twocolumn,authoryear]{elsarticle}

 \usepackage{amssymb}
 \usepackage{lipsum}
\usepackage{amsmath}
\usepackage{graphicx}
\usepackage{caption}
\usepackage{subcaption}
\usepackage{tikz}
\usepackage{bbm}
\usepackage{ulem}
\usepackage{algorithm}
\usepackage{algpseudocode}
\usepackage{mathtools}
\usepackage{hyperref}
\usetikzlibrary{shapes,arrows}
\DeclarePairedDelimiter\floor{\lfloor}{\rfloor}
\setcitestyle{numbers}
\DeclareMathOperator*{\argmax}{arg\,max}

\begin{document}

\begin{frontmatter}

\title{Study of  speaker localization with binaural microphone array incorporating auditory filters and lateral angle estimation}

\author[a]{Yanir Maymon\corref{cor1}}
\ead{yanirma@post.bgu.ac.il}
\author[b]{Israel Nelken}
\author[a]{Boaz Rafaely}

\cortext[cor1]{Corresponding author}

\address[a]{School of Electrical and Computer Engineering, Ben Gurion University of the Negev, Beer-Sheva 8410501, Israel}
\address[b]{Department of Neurobiology, The Hebrew University of Jerusalem, Jerusalem 9190401, Israel}

% 1
\begin{abstract}
Speaker localization for binaural microphone arrays has been widely studied for applications such as speech communication, video conferencing, and robot audition.
Many methods developed for this task, including the direct path dominance (DPD) test, share common stages in their processing, which include transformation using the short-time Fourier transform (STFT), and a direction of arrival (DOA) search that is based on the head related transfer function (HRTF) set.
In this paper, alternatives to these processing stages, motivated by human hearing, are proposed. These include incorporating an auditory filter bank to replace the STFT, and a new DOA search based on transformed HRTF as steering vectors.
A simulation study and an experimental study are conducted to validate the proposed alternatives, and both are applied to two binaural DOA estimation methods; the results show that the proposed method compares favorably with current methods. 
\end{abstract}

% 2
\end{frontmatter}
% 3
\section{Introduction}
Direction of arrival (DOA) estimation  of speakers in a room using a binaural array is a challenging problem which has a wide range of applications in speech enhancement, hearing aids and robot audition.
The challenge is exacerbated by coherent reflections that obscure DOA information typically available only in the direct sound.
Methods for DOA estimation using binaural arrays have been based on widely-used approaches for source localization. These include estimation of the interaural time difference (ITD), which is extracted from the generalized cross correlation (GCC) \cite{GCC}, beamforming based methods \cite{BMFM}, and subspace methods such as multiple signal classification (MUSIC) \cite{BMFM}.

In the last decades, new studies proposed techniques to make these methods more robust to reverberation. For example, a GCC based approach that is robust to noise and multipath distortion  for ITD estimation \cite{RGCC}, and the coherent signal subspace method (CSSM) \cite{CSSM}, which implements focusing and frequency smoothing in order to decorrelate coherent sources. 
Additionally, another GCC-based approach that employs a Bayesian framework has been developed. This approach utilizes a mixture model along with Bayesian modeling to robustly estimate the directions of multiple speakers in the presence of noise and reverberation \cite{escolano2014bayesian}. Furthermore, this Bayesian methodology has also been extended to more complex microphone arrays, such as coprime arrays \cite{bush2018model}, and spherical arrays \cite{landschoot2019model}. 
Recently, a reverberation-robust method, based on the CSSM and originally developed for spherical arrays, has been proposed, called the direct path dominance (DPD) test \cite{SDPD}. The estimation of the DOA is performed by selecting time-frequency (TF) bins that are dominated by the direct sound, thus successfully overcoming the detrimental effect of room reverberation. More recently, an extension for the DPD test for arbitrary arrays, and particularly for binaural arrays, was proposed \cite{ADPD,BDPD}.
This extension incorporates a
focusing process that does not require an initial DOA estimation, making it usable for reverberant environments.

The methods described above are all based on explicit processing of array data such as correlation matrices.
Recently, a deep neural network based methods has been developed for sound source localization in general \cite{grumiaux2022survey}, and binaural localization in particular \cite{P2018-DEEP,P2019,P2017}, offering new opportunities for exploiting information in the data. Unlike the DPD, these methods require a full learning phase with labeled data, which makes these methods less appropriate for some applications.

In summary, the methods presented above for DOA estimation using a binaural array, although showing good performance in many cases, may have limited performance for challenging environments with noise and reverberation. In particular, this paper examines two features of many current methods. First, current methods are mostly based on pre-processing using the fast Fourier transform (FFT). This is in contrast to the human ear, for example, that has filters whose bandwidth increases approximately proportionally to their center frequency \cite{AUDFB}. Second, many current methods \cite{DP-RTF,P2012,P2019,P2017}, simplify the directional search space, and assume, for example, that the sound source is positioned in the horizontal plane (elevation of $0^{\circ}$), relative to the binaural array, and then only search directly for the source azimuth angle.

In this paper, we develop and investigate alternatives to these commonly used processing features, showing that improved performance can indeed be achieved. The proposed alternatives can be integrated into many of the sate-of-the-art methods presented in this literature review, including in the preprocessing stages of neural network based methods. First, we present a processing framework which allows the incorporation of complex-valued version of the auditory filter bank in the processing pipeline, replacing the FFT. The new framework which can be incorporated in a wide range of current methods, shows that improved DOA performance can be achieved in some cases. Second, a new method is proposed to directly estimate the lateral angle in an interaural coordinate system \cite{INCSYS}, by incorporating the characteristics of the cone of confusion \cite{coc_book}, showing improved performance over standard azimuth and elevation based DOA estimation. An experimental study examines the proposed processing alternatives relative to the current approaches, when applied to a binaural DPD test-based method.

% 4
\section{System model and DPD test}
\label{sec:DPDT}
This section briefly presents the system model assumed in this work, and the DPD-test for a binaural array according to \cite{ADPD}.  
While the DPD test-based method presented here can indeed extended to the localization of multiple speakers, for the sake of simplicity and focus, this paper primarily explores the case of a single speaker. 
It is important to note that the DPD test based method is incorporated in this paper as an example of a state-of-the-art algorithm that can be applied to a binaural microphone array. Nevertheless, the processing alternatives developed and investigated in this paper can also be applied to other current methods of binaural speaker localization
\cite{BS1,BS2,BS3,BS4}.

Consider a binaural array, and $L$ plane waves forming the sound field around the array. 
Among these waves, one can be a direct sound from the source, while the rest are reflections from the room walls. 
The binaural signal can be expressed in the time domain as follows:
\begin{equation}\label{eq:1}
\mathbf{p}(t)= \sum_{i=1}^{L}
\mathbf{h}(t,\psi_{i}) \circledast s_{i}(t) + \mathbf{n}(t),
\end{equation}
where $t$ is time, $\mathbf{p}(t) = [p_{l}(t),p_{r}(t)]^T$ is the left and right binaural signals, $s_{i}(t)$ is the $i$'th source signal, $\mathbf{h}(t,\psi_{i})$ is the impulse response corresponding to the $i$'th plane wave direction of $\psi_i = (\theta_i,\phi_i)$, where $\theta_i$ and $\phi_i$ are the elevation and the azimuth of the source, respectively, and $\circledast$ denotes convolution. $\mathbf{n}(t) = [n_{l}(t),n_{r}(t)]^T$ is additive sensor noise.

By employing the multiplicative transfer function (MTF) approximation \cite{MTF}, the binaural signals can be expressed in the short-time Fourier transform (STFT) domain as follows:
\begin{equation}\label{eq:2}
\mathbf{p}(\tau,\omega)= \mathbf{H}(\omega,\psi)\mathbf{s}(\tau,\omega) + \mathbf{n}(\tau,\omega),
\end{equation}
where $\tau$ and $\omega$ are the time frame and frequency indices, respectively. $\mathbf{p}(\tau,\omega) = [p_{l}(\tau,\omega),p_{r}(\tau,\omega)]^T$ is the left and right binaural signals, $\mathbf{H}(\omega,\psi) = [\mathbf{h}(\omega,\psi_{1}),...,\mathbf{h}(\omega,\psi_{L})]^T$ is the $2 \times L$ head related transfer function (HRTF) matrix, and $\mathbf{s}(\tau,\omega) = [s_{1}(\tau,\omega),...,s_{L}(\tau,\omega)]^T$ is the vector of source signals. $\mathbf{n}(\tau,\omega) = [n_{l}(\tau,\omega),n_{r}(\tau,\omega)]^T$ is additive sensor noise.

In order to estimate the DOA of the source representing the direct sound, the next stage is to estimate the spatial spectrum matrix in every TF bin, by  averaging  over a predefined  range in time and frequency. 
This stage requires a focusing process to eliminate the frequency dependence of the HRTF matrix within the specified averaging frequency range \cite{BDPD}. 
This focusing process, crucial for maintaining spatial information within the smoothed HRTF matrix, involves aligning the HRTF matrices within the averaging window to the HRTF matrix from the center frequency. The alignment is implemented using a focusing matrix $\mathbf{T}(\omega,\omega_0)$ that satisfies the following:
\begin{equation}\label{eq:FocusingAlignment}
\mathbf{T}(\omega,\omega_0)\mathbf{H}(\omega,\psi) = \mathbf{H}(\omega_0,\psi).
\end{equation}
The transformed binaural signal $\Tilde{\mathbf{p}}(\tau,\omega)$ is then obtained by multiplying the original binaural signal $\mathbf{p}(\tau,\omega)$ by the focusing matrix: 

\begin{equation}\label{eq:Focusing}
\begin{aligned}
\Tilde{\mathbf{p}}(\tau,\omega) & = \mathbf{T}(\omega,\omega_0)\mathbf{p}(\tau,\omega) \\
& = \mathbf{H}(\omega_0,\psi)\mathbf{s}(\tau,\omega) + \Tilde{\mathbf{n}}(\tau,\omega),
\end{aligned}
\end{equation}
where $\Tilde{\mathbf{n}}(\tau,\omega) = \mathbf{T}(\omega,\omega_0)\mathbf{n}(\tau,\omega)$ is the transformed noise.

After the focusing process, a smoothing operation is performed. A spatial spectrum matrix, $\mathbf{R}(\tau,\omega)$, is computed at each time-frequency bin:
\begin{equation}\label{eq:SpectrumMatrix}
\mathbf{R}(\tau,\omega)= E{[\Tilde{\mathbf{p}}(\tau,\omega)\Tilde{\mathbf{p}}^H(\tau,\omega)]},
\end{equation}
where $E[\cdot]$ denotes expectation.
This matrix is estimated by averaging $J_{\tau}$ and $J_{\omega}$ adjacent time frames and frequency bins, respectively:

%
%\begin{equation}\label{eq:4}
\begin{multline}\label{eq:4}
\hat{\mathbf{R}}(\tau,\omega)= \\ \frac{1}{J_{\tau}J_{\omega}}
\sum_{j_{\tau}=0}^{J_{\tau} - 1} \sum_{j_{\omega}=0}^{J_{\omega} - 1} \Tilde{\mathbf{p}}(\tau-j_{\tau},\omega-j_{\omega})\Tilde{\mathbf{p}}^H(\tau-j_{\tau},\omega-j_{\omega}).
\end{multline}
%\end{equation}
%

In the next stage, the singular-value decomposition (SVD) of $\hat{\mathbf{R}}(\tau,\omega)$ is computed at each TF bin, in order to find bins that pass the DPD test \cite{SDPD}. The SVD operation splits the spatial spectrum matrix into signal and noise subspaces. This partitioning of the data is a fundamental step towards applying subspace methods, particularly the MUSIC algorithm. The SVD of the matrix $\hat{\mathbf{R}}$ can be expressed as:

\begin{equation}
\hat{\mathbf{R}}=\mathbf{Q\Sigma Q^H}=
\left[ \begin{array}{ccc}
\mathbf{q_s} & \mathbf{q_n}
\end{array} \right]
\left[ \begin{array}{ccc}
\mathbf{\sigma_s} & 0 \\
0 & \mathbf{\sigma_n}
\end{array} \right]
\left[ \begin{array}{ccc}
\mathbf{q_s}^H \\
\mathbf{q_n}^H
\end{array} \right],
\end{equation}
where, $\mathbf{q_s}$ and $\mathbf{q_n}$ represent the signal and noise subspaces, respectively, and $\mathbf{\sigma_s}$ and $\mathbf{\sigma_n}$ denote their corresponding singular values. The DPD test is then applied as follows:

\begin{equation}\label{eq:DPDT}
\mathcal{D}=\left\{(\tau,\omega): \frac{\sigma_s(\hat{\mathbf{R}}(\tau,\omega))}{\sigma_n(\hat{\mathbf{R}}(\tau,\omega))}\geq \mathcal{TH} \right\},
\end{equation}
where $\sigma_s$ and $\sigma_n$ denote the largest and second largest singular values. $\mathcal{TH}$ is a threshold, chosen sufficiently larger than one to ensure that $\hat{\mathbf{R}}$ are dominated by a single source.

In the next stage, a MUSIC spectrum is calculated for every TF bin that passes the DPD test, i.e. for all $(\tau,\omega) \in\mathcal{D}$ the MUSIC spectrum is computed by
\begin{equation} \label{eq:MUSIC}
P(\psi)=\frac{1}{||\mathbf{q}_{n}^H \mathbf{{h}}(\psi)
||^2},
\end{equation}
where $\psi$ represents the direction on a two-dimensional (2D) search grid, and $\mathbf{{h}}(\psi)$ is the steering vector in the direction $\psi$.

The direction $\psi$ that maximizes the MUSIC spectrum is the DOA estimate for the specific bin.
This process leads to a DOA histogram containing all DOA estimates for all TF bins that passed the DPD test, denoted by the set $\left\{\psi_\mathcal{D}\right\}$.

In the final stage, the source direction can be estimated by taking the average angle of the DOA histogram or by performing clustering, as suggested in \cite{GMM}. In this paper we will use the first and simpler method,

\begin{equation} \label{eq:MeanAz}
\hat{\psi} = \overline{\left\{\psi_{\mathcal{D}}\right\}},
\end{equation}
where $\overline{\left\{\cdot\right\}}$ denotes the averaging operation.

\begin{figure}[htb]
\centering
% tarshim
%
\begin{center} 

\tikzstyle{block} = [draw, rectangle, 
    minimum height=2em, minimum width=3em]
\tikzstyle{meoyan} = [draw, diamond, aspect=2]
\tikzstyle{sum} = [draw, circle, node distance=1cm]
\tikzstyle{input} = [coordinate]
\tikzstyle{output} = [coordinate]
\tikzstyle{pinstyle} = [pin edge={to-,thin,black}]

% The block diagram code is probably more verbose than necessary
\begin{tikzpicture}[auto, node distance=1.5cm,>=latex']
    % We start by placing the blocks
    \node [input, name=input] {};
    \node [block, right of=input,align=center] (b1) {\scriptsize STFT};
    \node [block, right of=b1,
    node distance=2.5cm,align=center] (b2) {\scriptsize Focusing};
    \node [block, right of=b2,
    node distance=2.5cm,align=center] (b3) {\scriptsize TF \\[-0.3em] \scriptsize Averaging };
   
    % We draw an edge between the b1 and b2 block to 
    % calculate the coordinate u. We need it to place the measurement block. 
    \draw [->] (input) -- node [minimum height=1.2cm][left=0.5cm]{\scriptsize $\mathbf{p}(t)$} (b1);
    \draw [->] (b1) -- node[name=P2] [minimum height=1.2cm]{\scriptsize $\mathbf{p}(\tau,\omega)$} (b2);
    \draw [->] (b2) -- node[name=P3] [minimum height=1.2cm]{\scriptsize $\mathbf{\Tilde{p}}(\tau,\omega)$} (b3);
    
    \node [block, below of=b3,
   node distance=2cm,align=center] (b4) {\scriptsize SVD};
   \node [block, below of=b2,
   node distance=2cm,align=center] (b4_dpd) {\scriptsize DPD Test};
   \node [block, below of=b1,
   node distance=2cm,align=center] (b5) {\scriptsize MUSIC};
   \node [meoyan, below of=b5,
   node distance=1.5cm,align=center] (b5u) {\scriptsize HRTF};
   \node [output,left of=b5,name=output] {};

    \draw [->] (b3) -- node [minimum height=1.2cm]{\scriptsize $\mathbf{R}(\tau,\omega)$} (b4);
    \draw [->] (b4) -- node [minimum height=1.2cm]{\scriptsize} (b4_dpd);
    \draw [->] (b4_dpd) -- node [minimum height=1.5cm] [above=0.1cm] {\scriptsize $\{\mathbf{q}_{n}(\tau,\omega)\}_{(\tau,\omega) \in \mathcal{D}}$} (b5);
    \draw [->] (b5) -- node [minimum height=1.2cm] [left=0.5cm] {\scriptsize {$\hat{\psi}$}} (output);
    \draw [->] (b5u) -- node [minimum height=1.2cm] {\scriptsize $\mathbf{h}(\psi,\omega)$} (b5);
           
    %\node [output, right of=b4] (output) {};
\end{tikzpicture}
\end{center}
\caption{Block diagram of the DPD algorithm for DOA estimation}
\label{fig:block1}
\end{figure}
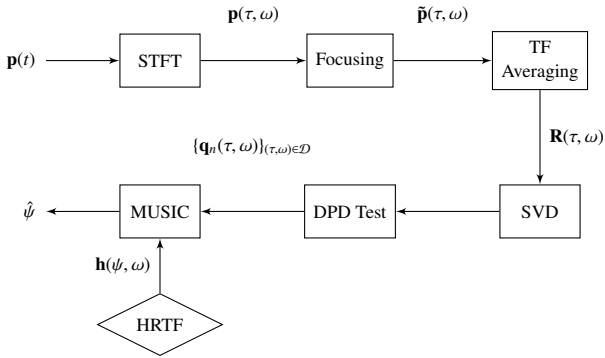

Fig. \ref{fig:block1} shows a block diagram of the DPD test based algorithm used to estimate the direction angle.

%

% 5
\section{Replacing FFT with auditory filters}
In this section we present the first proposed innovation in the processing pipeline: developing a formulation to enable the incorporation of human-hearing motivated auditory filter banks to replace the FFT. 

The following formulation is developed for the auditory filter, so that it can replace the FFT. Note that auditory filters are typically employed to compute signal power in frequency bands, while here a version of the filtering is required that provides both magnitude and phase similar to the FFT. 

\begin{figure*}[ht]
     \centering
     \begin{subfigure}[b]{0.45\textwidth}
         \centering
         \includegraphics[width=\textwidth]{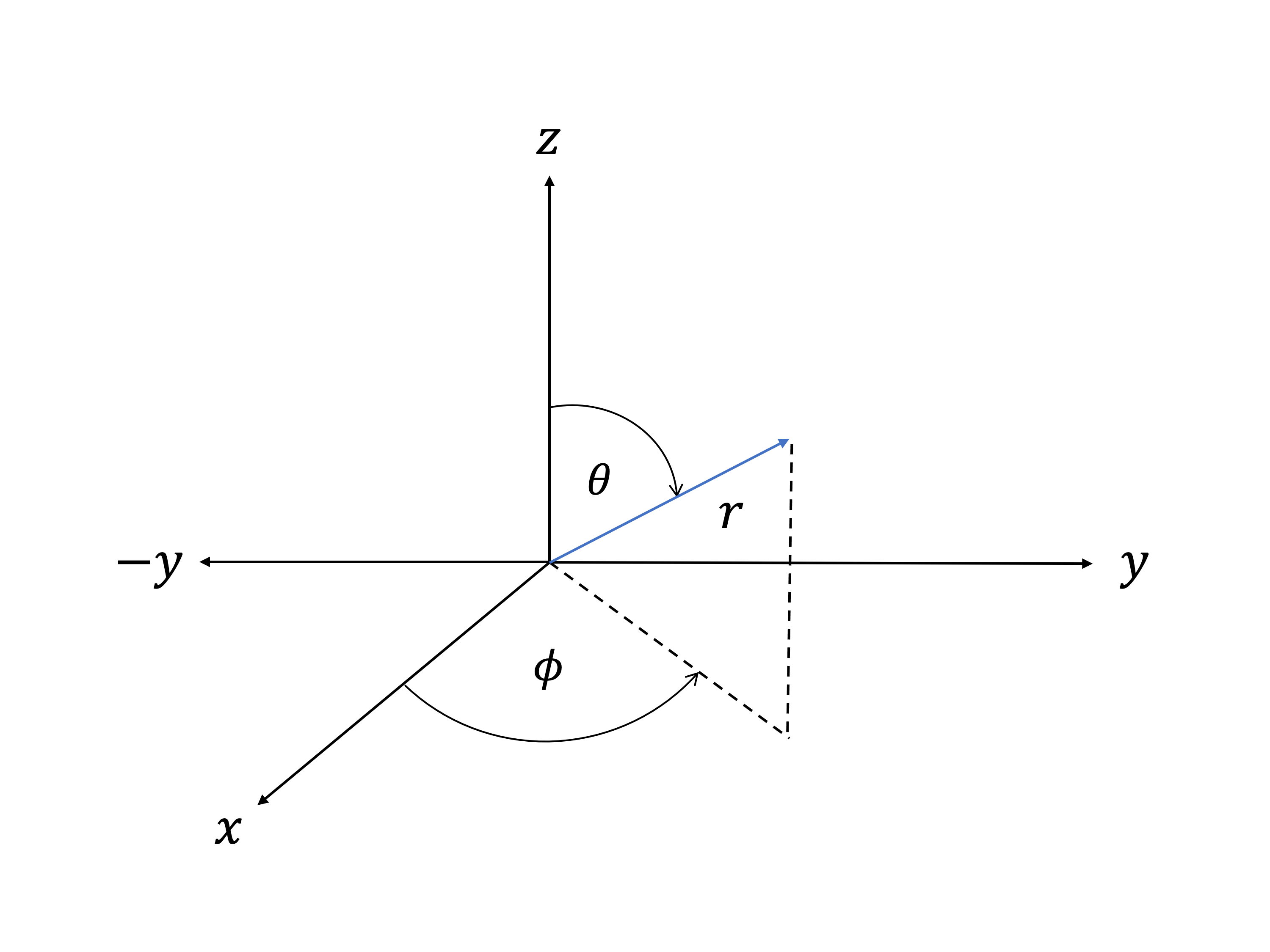}
         \caption{}
     \end{subfigure}
     \hfill
     \begin{subfigure}[b]{0.45\textwidth}
         \centering
         \includegraphics[width=\textwidth]{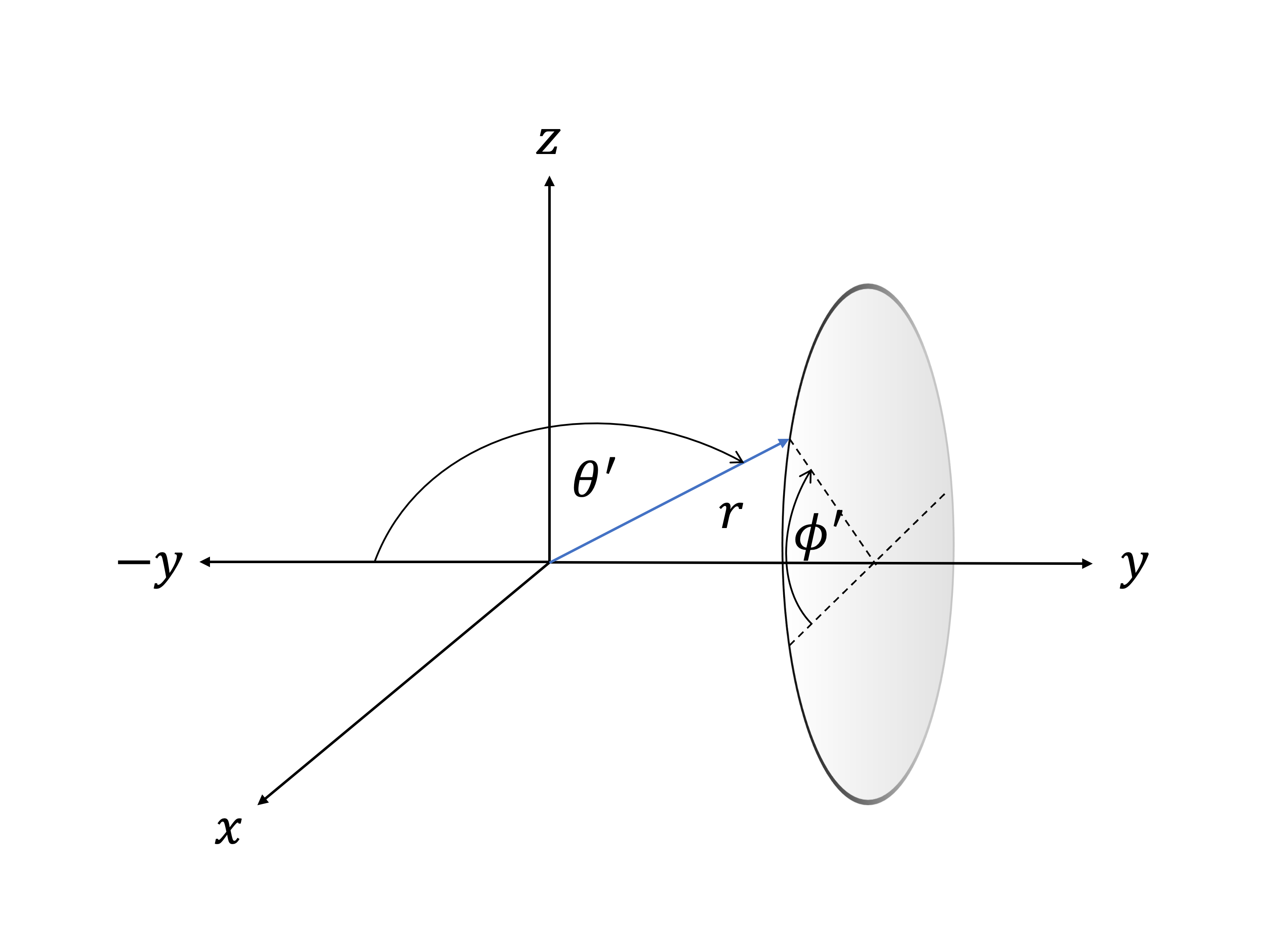}
         \caption{}
     \end{subfigure}
     \hfill
    \caption{ (a) Spherical coordinate system and (b) Interaural coordinate system, both presented over the Cartesian coordinate system }
    \label{fig:coordinates}
\end{figure*}

The starting point is the continuous time STFT, or the windowed Fourier transform (WFT) defined as \cite{STFT}
\begin{equation}
\label{eq:stft}
X(\tau,\omega_{c}) = \int_{-\infty}^{\infty} x(t)w(t-\tau)e^{-j\omega_{c} t} dt,
\end{equation}
where $x(t)$ is an arbitrary signal and $w(t)$ is a window function centered around zero.
The above equation can be reinterpreted in a form that characterizes a signal passing through a filter. This can be expressed as \cite{STFT_FILTER}

\begin{equation}
\label{eq:stft2}
\begin{aligned}
X(\tau,\omega_{c})  = e^{-j\omega_{c} \tau} 
\int_{-\infty}^{\infty} x(t)f(\tau - t ;\omega_{c}) dt,
\end{aligned}
\end{equation}

where \eqref{eq:stft2} denotes the convolution between the signal $x(t)$ and a filter $f(t;\omega_{c})$. The filter $f(t;\omega_{c})$ is defined as: \color{black}
\begin{equation}
\label{eq:filtc}
f(t;\omega_{c}) = w(-t) e^{j \omega_{c} t}.
\end{equation}
Note that \eqref{eq:filtc} can be considered as a one-sided version (over the frequency axis) of the filter defined by the window function $w(t)$ centered around $\omega_{c}$. Next, \eqref{eq:stft2}  can be formulated as
\begin{equation}
\label{eq:stftfreq}
\begin{aligned}
X(\tau,\omega_{c}) &  = e^{-j\omega_{c} \tau} \mathcal{F}^{-1} \{X(\omega)F(\omega;\omega_{c})\}(\tau),
\end{aligned}
\end{equation}
where $\mathcal{F}^{-1}$ denotes the inverse Fourier transform, and $X(\omega)$ and $F(\omega;\omega_{c})$ are the Fourier transforms of $x(t)$ and $f(t;\omega_{c})$ respectively.
Overall, \eqref{eq:stftfreq} can be interpreted as filtering with a one sided band-pass filter, centered around the positive frequency $\omega_{c}$, and then shifting the output signal back to baseband to be centered around the origin.

This general formulation allows the incorporation of auditory filter banks to replace the conventional FFT. For this, the filter $F(\omega;\omega_{c})$ is replaced in this work with a gammatone filter bank, which is designed to model the human auditory system. The filter has been specifically modified to be one-sided in the frequency domain. 

The gammatone filters' impulse response is defined as \cite{patterson1992complex}
\begin{equation}
\label{eqn:gt}
gt(t;\omega_{c}) = 
\begin{cases}
t^{n-1}e^{-2\pi b_{\omega_{c}} t}\cos(\omega_{c} t), & \text{if}\ t \geq 0 \\
0, & \text{otherwise},
\end{cases}
\end{equation}
where $\omega_{c}$ is the center frequency of the filter in channel $c$, $n$ is the filter order, which is set to be 4 in this work, and $b_{\omega_{c}}$ is the filter bandwidth.
The center frequencies and bandwidths of the filters are determined according to the Equivalent
Rectangular Bandwidth (ERB) scale \cite{AUDFB}.

By keeping the envelope of the gammatone impulse response, a one sided gammatone filter can be constructed as follows:

\begin{equation}
\label{eq:gtf}
g(t;\omega_{c}) = w_{gt}(-t;\omega_{c}) e^{j \omega_{c} t},
\end{equation}
where
\begin{equation}
\label{eqn:gt15}
w_{gt}(t;\omega_{c}) = 
\begin{cases}
t^{n-1}e^{-2\pi b_{\omega_{c}} t}, & \text{if}\ t \geq 0 \\
0, & \text{otherwise},
\end{cases}
\end{equation}
Eq. \eqref{eq:stftfreq} can now be rewritten for the case of the gammatone  filters as follows:
\begin{equation}
\label{eq:gtfb_f}
X_{AFB}(\tau,\omega_{c})  = e^{-j\omega_{c} \tau} \mathcal{F}^{-1} \{X(\omega)G(\omega;\omega_{c})\}(\tau),
\end{equation}
where $G(\omega;\omega_{c})$ is the Fourier transform of $g(t;\omega_{c})$ defined in Eq. \eqref{eq:gtf}, and the subscripts AFB is acronyms of Auditory Filter Bank, representing the alternative operation presented here.

The above formulation is developed in continuous time. Practically, the signals are discrete, and a few modifications are required to adapt the above formulation to a discrete-time signal. 
Equation \eqref{eq:stft2} can be rewritten for the discrete-time case as follows:
\begin{equation}
\label{eq:discrete_stft}
X[m,k_{c}]  = e^{-j2\pi \frac{k_{c}}{N}m}\sum_{n=0}^{N-1} x[n]f[m-n],
\end{equation}
where $x[n]$ is the arbitrary signal of length $N$, and $f[n]$ is the discrete version of the filter defined in Eq. \eqref{eq:filtc}, of length $L$. In order to formulate Eq. \eqref{eq:discrete_stft} using FFT, zero padding on the signals to length $N + L - 1$ is required. Define the zero padding signals of $x[n]$ and $f[n]$ as $\Tilde{x}[n]$ and $\Tilde{f}[n]$, respectively.
Following that, Eq. \eqref{eq:gtfb_f} can be rewritten as
\begin{equation} \label{eq:discrete_af}
X_{AFB}[m,k_{c}]  = e^{-j2\pi \frac{k_{c}}{N}m} FFT^{-1} \{\Tilde{X}[k]\Tilde{G}[k;k_{c}]\}[m],
\end{equation}
where $\Tilde{X}[k]$ and $\Tilde{G}[k;k_{c}]$ are the FFT of $\Tilde{x}[n]$ and $\Tilde{g}[n;k_{c}]$, where $\Tilde{g}[n;k_{c}]$ is defined using the discrete version of Eq. \eqref{eq:gtf}, with zero padding.

Sampling of the auditory filter output signals can be performed in a similar way to the STFT; however, unlike the STFT, each channel has to be sampled at different time intervals, because each channel has a different bandwidth determined by the ERB. Hence, according to the Nyquist sampling theorem, the sampling time interval can be defined as follows \cite{Nyq}:
\begin{equation}\label{eq:tauerb}
\Delta \tau(k_{c}) = \frac{1}{2BW_{ERB}(k_{c})}
\end{equation},
where $BW_{ERB}(k_{c})$ is the bandwidth in the channel $c$.
For the discrete case, sampling is replaced by decimating the signal at the filter output. The decimation factor can be computed as follows:
\begin{equation}\label{eq:tauerb2}
M(k_{c}) =  \floor*{\frac{\Delta \tau(k_{c})}{T_{s}}},
\end{equation}
where $T_{s}$ is the sampling interval of the continuous time signal.
Now, the filters' output can be computed as follows:
\begin{equation}\label{eq:tauerb3}
\Tilde{X}_{AFB}[m,k_{c}] = {X}_{AFB}[mM(k_{c}),k_{c}].
\end{equation}

% 6
\section{Lateral angle estimation}

\begin{algorithm}[h]
\caption{Binaural Direction-of-Arrival (DOA) Estimation}
\label{alg:doa_estimation}
\begin{algorithmic}[1]

\Statex \textbf{Input:} The binaural signal $\mathbf{p}(t)$
\Statex \textbf{Output:} Estimated lateral angle $\hat{\theta^{'}}$
\Statex \textbf{Required:} Lateral steering vectors set $\mathbf{u}_{1}(\theta^{'})$
\State $\begin{aligned}[t]
                 &\text{Apply AFB to the binaural signal} \\[-0.5ex]
                 & \mathbf{p}_{AFB}(\tau, \omega) = \text{AFB}(\mathbf{p}(t))
                \end{aligned}$
\State $\begin{aligned}[t]
         &\text{Multiply the binaural signal by the focusing matrix}\\[-0.5ex]
         &\mathbf{\Tilde{p}}_{AFB}(\tau,\omega) = \mathbf{T(\omega, \omega_{0})} \mathbf{p}_{AFB}(\tau, \omega)
        \end{aligned}$
\State $\begin{aligned}[t]
         &\text{Estimate the spatial spectrum matrix}\\[-0.5ex]
         &\hat{\mathbf{R}}(\tau,\omega)= \\
         & \frac{1}{J_{\tau}J_{\omega}}
         \sum_{j_{\tau}=0}^{J_{\tau} - 1} \sum_{j_{\omega}=0}^{J_{\omega} - 1} \mathbf{\Tilde{p} }_{AFB}(\tau-j_{\tau},\omega-j_{\omega})\mathbf{\Tilde{p}}_{AFB}^H(\tau-j_{\tau}, \omega-j_{\omega}).
        \end{aligned}$
\State $\begin{aligned}[t]
         &\text{Compute the eigenvalue ratio and perform the DPD test}\\[-0.5ex]
         &\mathcal{D}=\left\{(\tau,\omega): \frac{\sigma_s(\hat{\mathbf{R}}(\tau,\omega))}{\sigma_n(\hat{ \mathbf{R}}(\tau,\omega))}\geq \mathcal{TH} \right\}
        \end{aligned}$
\For{each $i \in \mathcal{D}$}
     \State$\begin{aligned}[t]
         &\text{Compute the MUSIC spectrum}\\[-0.5ex]
         & P_{i}(\theta^{'})=\frac{1}{||\mathbf{q}_{n,i}^H \mathbf{u}_1(\theta^{'})| |^2}
        \end{aligned}$
     \State$\begin{aligned}[t]
         &\text{Estimate the lateral angle}\\[-0.5ex]
         &\hat{\theta^{'}}_{i} = \argmax_{\theta^{'}} P_{i}(\theta^{'})
        \end{aligned}$
\EndFor
\State$\begin{aligned}[t]
     &\text{Compute the average lateral angle based on all estimated} \\[-0.5ex]
     &\text{angles}\\[-0.5ex]
     & \hat{\theta^{'}} = \frac{1}{|\mathcal{D}|} \sum_{i \in \mathcal{D}} \hat{\theta^{'}}_{i}
\end{aligned}$
\Statex \textbf{Return:} $\hat{\theta^{'}}$
\end{algorithmic}
\end{algorithm}

The previous section incorporated auditory filter banks in the processing,  which were motivated by human hearing. On a similar note, this section incorporates novel lateral angle estimation, also partially motivated by human hearing.
Binaural cues contain important  information about the azimuth direction of a source and its estimation is often a major goal in source localization. There are a number of options for estimating source azimuth given binaural steering vectors. The first is to perform a full 2D search over both azimuth and elevation, and then extract only the azimuth angle. This option may be computationally expensive due to the extensive search over all directions. Furthermore, this approach may suffer from error due to the less informative elevation cue \cite{EleVsAz}. Another option is to perform a one-dimensional (1D) search for source's azimuth by assuming the sources elevation is known, e.g., assuming sources are in the horizontal plane \cite{DP-RTF,P2012,P2019,P2017}. This option may be prone to error if the source's elevation is not accurately provided.

In this section a new localization framework is presented, aiming to overcome the limitations of previous approaches.

The proposed approach is motivated by the human auditory system, which relies on the ITD and the interaural level difference (ILD) as localization cues.  The set of source directions with a similar ITD and ILD form a cone, which is known as the cone of confusion \cite{coc_book}. Therefore, interaural cues can be used to distinguish between sources at different cones, but not between sources within the same cone.
Therefore, the interaural coordinate system, which directly represents the cone of confusion, may be more suitable than the standard spherical polar system. The spherical and interaural coordinate systems are illustrated in Fig. \ref{fig:coordinates}. 
Within the interaural coordinate system, the lateral angle differentiates between cones, and the intraconic angle indicates the position within a cone. 

Inspired by human localization, as discussed above, a new method for directly estimating the lateral angle is developed in this section.
The HRTF set, which is usually sampled in a spherical coordinates grid (azimuth and elevation), is resampled into a lateral-intraconic grid.
For each lateral angle in the set, a steering matrix is reconstructed with steering vectors representing all intraconic directions. This steering matrix of size $2 \times N$ is defined as
\begin{equation} \label{eq:HRTFMatrix}
\mathbf{H}(\theta^{'},\omega) \triangleq	
\left[ \begin{array}{cccc}
h_{1l}(\theta^{'},\omega)  & h_{2l}(\theta^{'},\omega) & ... &
h_{Nl}(\theta^{'},\omega) \\
h_{1r}(\theta^{'},\omega)  & h_{2r}(\theta^{'},\omega) & ... &
h_{Nr}(\theta^{'},\omega) 
\end{array} \right],
\end{equation}
where $\theta^{'}$ is the lateral angle, $N$ is the number of the intraconic directions, and $h_{il}(\theta^{'},\omega)$ and $h_{ir}(\theta^{'},\omega)$ are the left and right HRTFs, respectively, where the subscripts $r$, $l$ and $i$ denote left, right and intraconic number in the set, respectively.
The columns are expected to be similar, but not the same, due to the HRTF similarity within a cone. 
We aim to find a single steering vector that best represents a specific lateral direction, i.e. a single steering vector for every cone. To do that, we decompose matrix $\mathbf{H}(\theta^{'},\omega)$ using SVD, as follows:
\begin{equation}
\mathbf{H}(\theta^{'},\omega) =
\mathbf{U}(\theta^{'},\omega) \mathbf{S}(\theta^{'},\omega) \mathbf{V}(\theta^{'},\omega),
\end{equation}
where the matrix $\mathbf{U}(\theta^{'},\omega)$ has the following structure:
\begin{equation} \label{eq:LatSV}
\mathbf{U}(\theta^{'},\omega) = \left[ \begin{array}{cccc}
\mathbf{u}_1(\theta^{'},\omega)  & \mathbf{u}_2(\theta^{'},\omega)  
\end{array} \right].
\end{equation}
The first column of $\mathbf{U}(\theta^{'},\omega)$, $\mathbf{u}_1(\theta^{'},\omega)$, corresponding to the largest singular value, is the best representation of a lateral direction that is common to all intraconic directions.
Therefore, we can use vector $\mathbf{u}_1(\theta^{'},\omega)$ as a steering vector for a 1D lateral search.

Then, similarly to in Eq. \eqref{eq:MUSIC}, the MUSIC spectrum will be calculated as follows:
\begin{equation} \label{eq:MUSICLAT}
P(\theta^{'})=\frac{1}{||\mathbf{q}_{n}^H \mathbf{u}_1(\theta^{'})||^2},
\end{equation}
where $\theta^{'}$ represents the lateral direction in a 1D grid, and $\mathbf{u}_1(\theta^{'})$ is the lateral steering vector (the time-frequency dependence is omitted for simplicity).

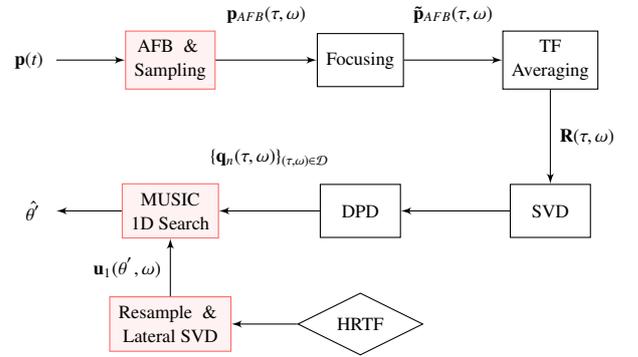
\begin{figure}[htb] 
\centering
% tarshim
%
\begin{center} 

\tikzstyle{block} = [draw, rectangle, 
    minimum height=2em, minimum width=3em]
\tikzstyle{new_block} = [draw, rectangle, 
    minimum height=2em, minimum width=3em,draw=red!60, fill=red!5]
\tikzstyle{meoyan_b} = [draw, diamond, aspect=2]
\tikzstyle{input} = [coordinate]
\tikzstyle{output} = [coordinate]

% The block diagram code is probably more verbose than necessary
\begin{tikzpicture}[auto, node distance=1.5cm,>=latex']
    % We start by placing the blocks
    \node [input, name=input] {};
    \node [new_block, right of=input,align=center] (b1) {\scriptsize AFB $\And$ \\[-0.3em] \scriptsize Sampling };
    \node [block, right of=b1,
    node distance=2.5cm,align=center] (b2) {\scriptsize Focusing};
    \node [block, right of=b2,
    node distance=2.5cm,align=center] (b3) {\scriptsize TF \\[-0.3em] \scriptsize Averaging };
   
    % We draw an edge between the b1 and b2 block to 
    % calculate the coordinate u. We need it to place the measurement block. 
    \draw [->] (input) -- node [minimum height=1.2cm][left=0.5cm]{\scriptsize $\mathbf{p}(t)$} (b1);
    \draw [->] (b1) -- node[name=P2] [minimum height=1.2cm]{\scriptsize $\mathbf{p}_{AFB}(\tau,\omega)$} (b2);
    \draw [->] (b2) -- node[name=P3] [minimum height=1.2cm]{\scriptsize $\mathbf{\Tilde{p}}_{AFB}(\tau,\omega)$} (b3);
    
    \node [block, below of=b3,
   node distance=2cm,align=center] (b4) {\scriptsize SVD};
   \node [block, below of=b2,
   node distance=2cm,align=center] (b4_dpd) {\scriptsize DPD};
   \node [new_block, below of=b1,
   node distance=2cm,align=center] (b5) {\scriptsize MUSIC \\[-0.3em] \scriptsize 1D Search};
   
    \draw [->] (b3) -- node[name=P4] [minimum height=1.2cm]{\scriptsize $\mathbf{R}(\tau,\omega)$} (b4);
    \draw [->] (b4) -- node[name=P4] [minimum height=1.2cm]{\scriptsize} (b4_dpd);
    \draw [->] (b4_dpd) -- node [minimum height=1.2cm] [above=0.1cm]{\scriptsize $\{\mathbf{q}_{n}(\tau,\omega)\}_{(\tau,\omega) \in \mathcal{D}}$} (b5);
    \draw [->] (b5) -- node [minimum height=1.2cm] [left=0.5cm] {\scriptsize {$\hat{\theta^{'}}$}} (output);
     
   \node [meoyan_b, below of=b4_dpd,
   node distance=1.5cm,align=center] (b4u) {\scriptsize HRTF};
   \node [new_block, below of=b5,
   node distance=1.5cm,align=center] (b5u) {\scriptsize Resample $\And$ \\[-0.3em] \scriptsize Lateral SVD };
   
   \draw [->] (b4u) -- node [minimum height=1.2cm] {} (b5u);
   \draw [->] (b5u) -- node [minimum height=1.2cm] {\scriptsize $ \mathbf{u}_{1}(\theta^{'},\omega)$} (b5);

   \node [output,left of=b5,name=output] {};      
    %\node [output, right of=b4] (output) {};
\end{tikzpicture}
\end{center}
\caption{Block diagram of the updated DPD algorithm with the incorporation of the auditory filter bank and the direct lateral angle estimation}
\label{fig:block3}
\end{figure}

Figure \ref{fig:block3} shows a block diagram of the DPD algorithm with the proposed auditory processing and with the incorporation of the auditory filter banks. 
The localization process is summarized in Algorithm \ref{alg:doa_estimation}.

\begin{figure*}[h!]
     \centering
     \begin{subfigure}[b]{0.45\textwidth}
         \centering
         \includegraphics[width=\textwidth]{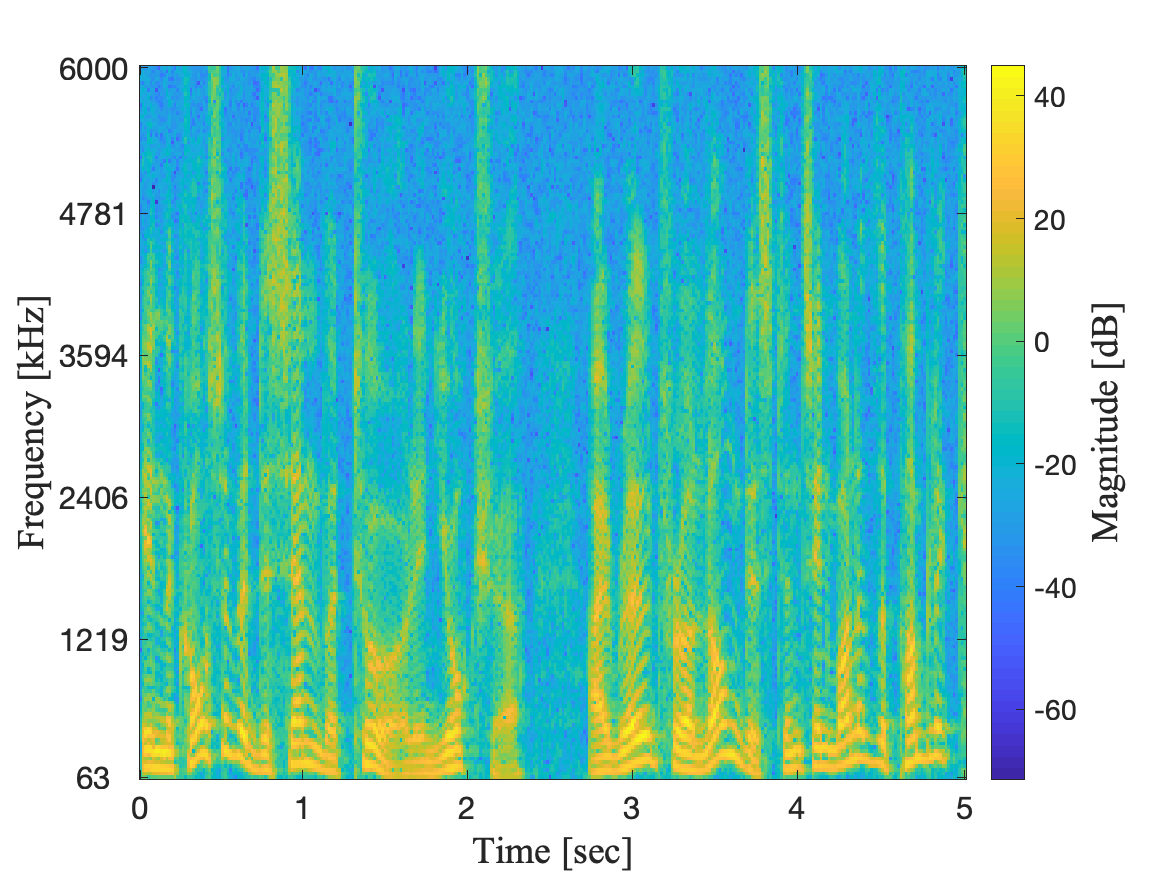}
         \caption{}
     \end{subfigure}
     \hfill
     \begin{subfigure}[b]{0.45\textwidth}
         \centering
         \includegraphics[width=\textwidth]{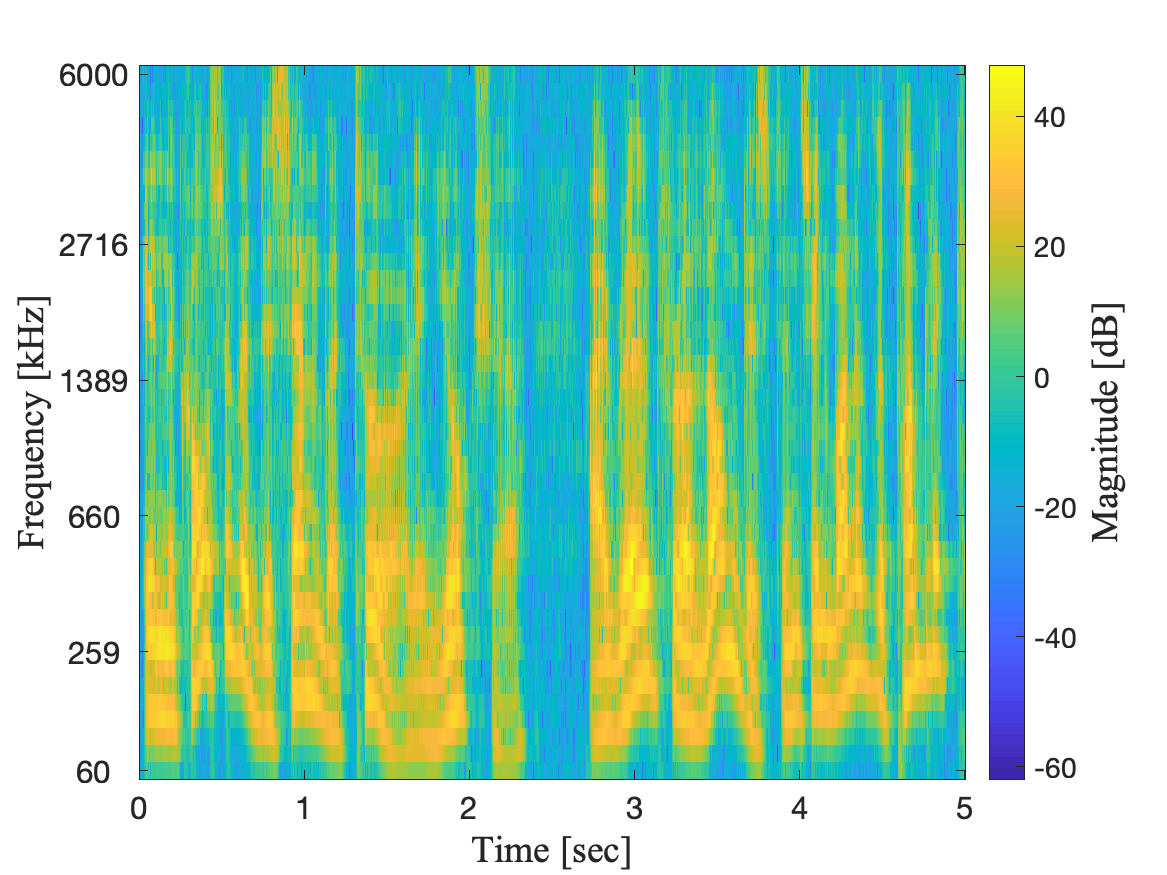}
         \caption{}
     \end{subfigure}
     \hfill
    \caption{(a) STFT magnitude of clean speech (b) AFB magnitude of clean speech}
    \label{fig:STFT}
\end{figure*}

\section{Simulation study}\label{sec:Study}

This section studies,through simulations, the performance of two DOA estimation methods for binaural arrays, and compares the results to those obtained when incorporating the proposed auditory processing,
under different reverberation and background noise conditions. 
The effect of the auditory filter bank and the direct lateral search, in terms of accuracy of DOA estimation, are investigated. 
The selected methods for comparison include the DPD method \cite{ADPD}, as presented in this paper, and  the joint estimation (JE) method proposed by Raspaud \cite{JEM_R}. These were selected as examples of methods that incorporate STFT computation and angle estimation searches based on HRTFs.
\subsection{Simulation setup}\label{sec:Setup}
The simulation setup includes a single speaker in a room, recorded by a binaural array. The room is rectangular, with various different dimensions, the speaker is represented by a point source, and the binaural arrays are simulated using a model of  HRTFs from the Neumann KU-100 manikin \cite{HRTF}.

In order to calculate the microphone signals, the room impulse responses are first computed using the image method \cite{ImMeth}, and then convolved with a speech signal of duration about $5\,$s, sampled at $48\,$kHz. After calculating the binaural signal, a Gaussian white noise source is added, in the form of sensor noise.

As there are no agreed benchmarks for binaural localization, and performance varies greatly based on the environment, we will run 500 simulations under different conditions so that a diverse set of conditions is generated, in order to obtain representative results. In each simulation, the room size, the speakers and the distance between the speaker and array is chosen randomly from a defined set of options. The room size options are 5 $\times$ 10 $\times$ 8 m, 9 $\times$ 7 $\times$ 5 m and 8 $\times$ 5 $\times$ 3 m, the speakers set, taken from
the TIMIT database \cite{TIMIT}, includes 2 male and 2 female speakers, and the distance options between the speaker and array are 0.5, 1 and 2 times the critical distance. The array position is chosen randomly within the boundaries of the room, and the speaker position is determined by the DOA, chosen randomly, and by the speaker-array distance.

Several options of signal to noise ratios (SNRs) and reverberation times ($T_{60}$) are generated. The SNR options go from very low value, -5 dB to
high value, 15 dB, with steps of 5 dB, and the $T_{60}$ options go from medium
reverberation time, $0.4$\,s to high reverberation time $0.8$\,s, with steps of $0.2$\,s.

% The Numman KU-100 \cite{HRTF} is used in this paper for the binaural array. The simulation includes a single speaker in a room.  We generate three rooms configurations with the sizes 15 $\times$ 10 $\times$ 8 $[m]$, 9 $\times$ 7 $\times$ 5 $[m]$ and 8 $\times$ 5 $\times$ 3 $[m]$. The reflection coefficients of the wall surface are the same over frequency and range from 0.6 to 0.9 with step of 0.1. The speaker-receiver distances are 0.5, 1 and 2, times the critical distance.
% From the above, the reverberation time is range from 0.1 to 1 $[sec]$.
% To calculate the microphone signals, the room impulse responses are first computed using the image method \cite{ImMeth}, and then convolved with speech of a duration of about $5\,$s, taken from the TIMIT database \cite{TIMIT}, and sampled at $48\,$kHz. The speakers selected from set of 4 speakers, 2 female and 2 male. Background white noise with a signal-to-noise ratio that range from -5 to 15, with step of 5,  was added, in the form of a diffuse sound field.

\subsection{Methodology}\label{sec:Methodology}

\begin{figure*}[h!]
     \centering
     \begin{subfigure}[b]{0.33\textwidth}
         \centering
         \includegraphics[width=\textwidth]{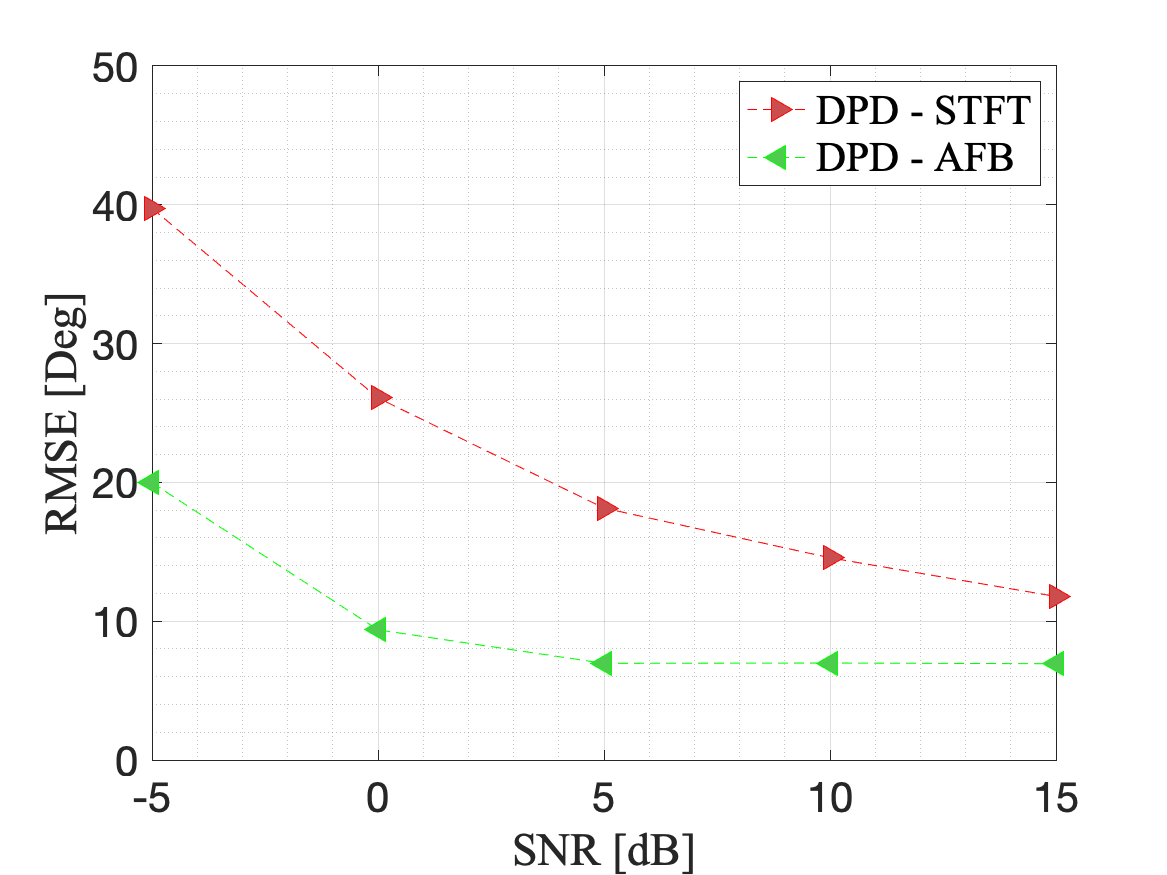}
         \caption{}
     \end{subfigure}
     \hfill
     \begin{subfigure}[b]{0.33\textwidth}
        \centering
        \includegraphics[width=\textwidth]{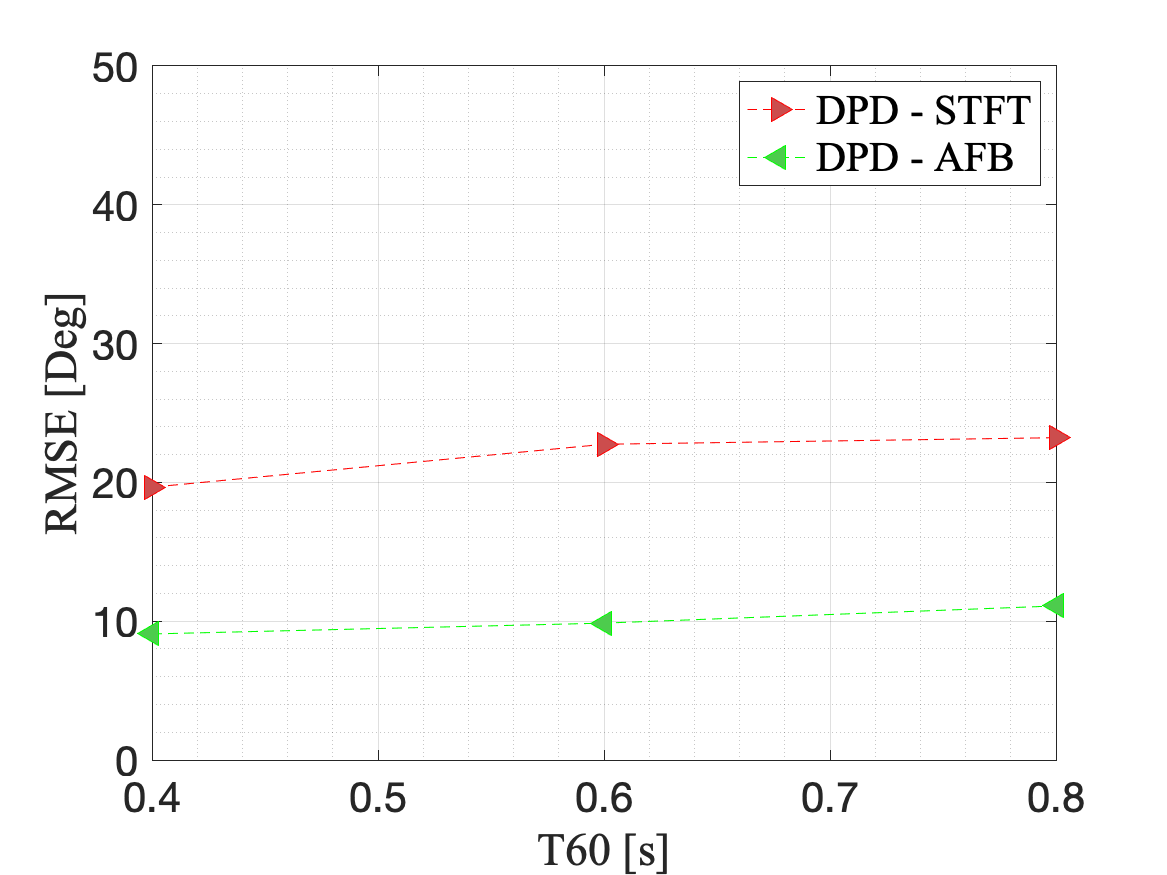}
        \caption{}
     \end{subfigure}
     \hfill
     \begin{subfigure}[b]{0.33\textwidth}
         \centering
         \includegraphics[width=\textwidth]{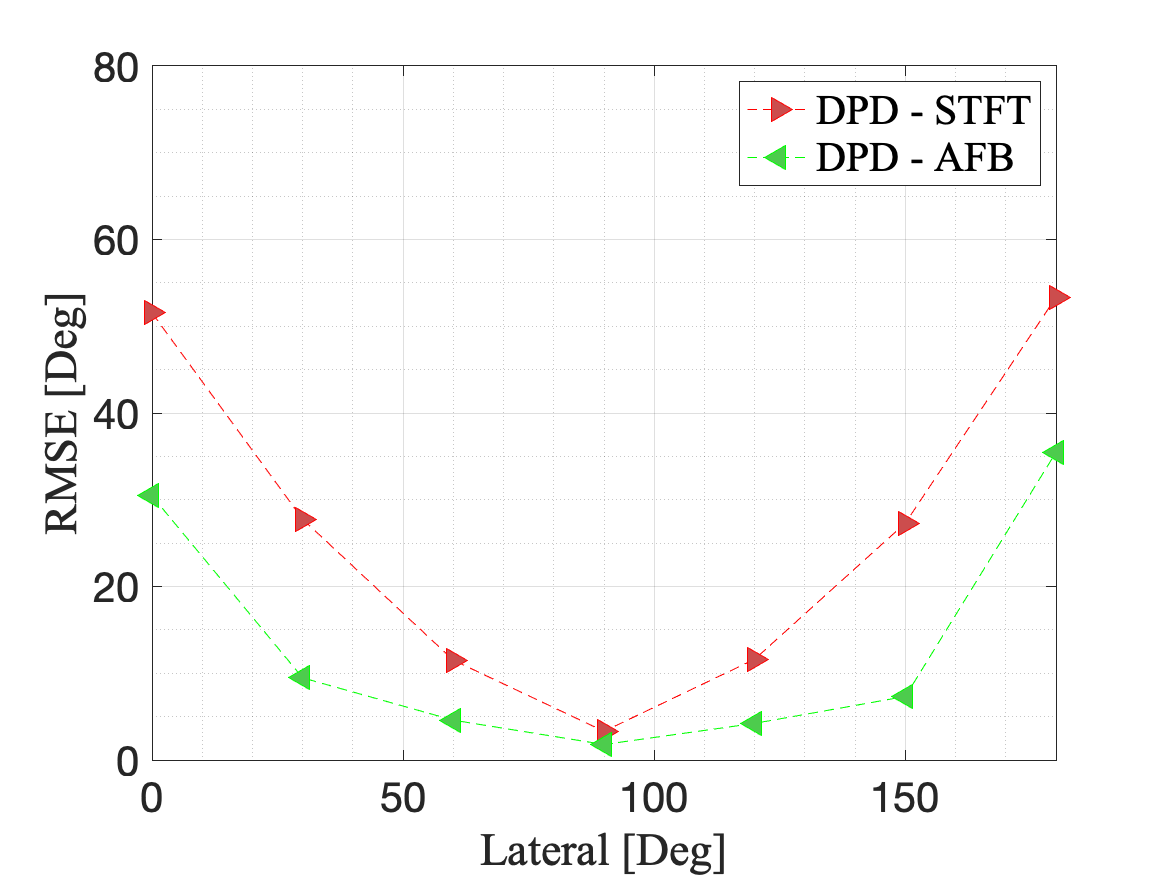}
         \caption{}
         \label{fig:DPD_JE_STFT2D_AUDFB2D_c}
     \end{subfigure}
     \vskip\baselineskip
     \begin{subfigure}[b]{0.33\textwidth}
         \centering
         \includegraphics[width=\textwidth]{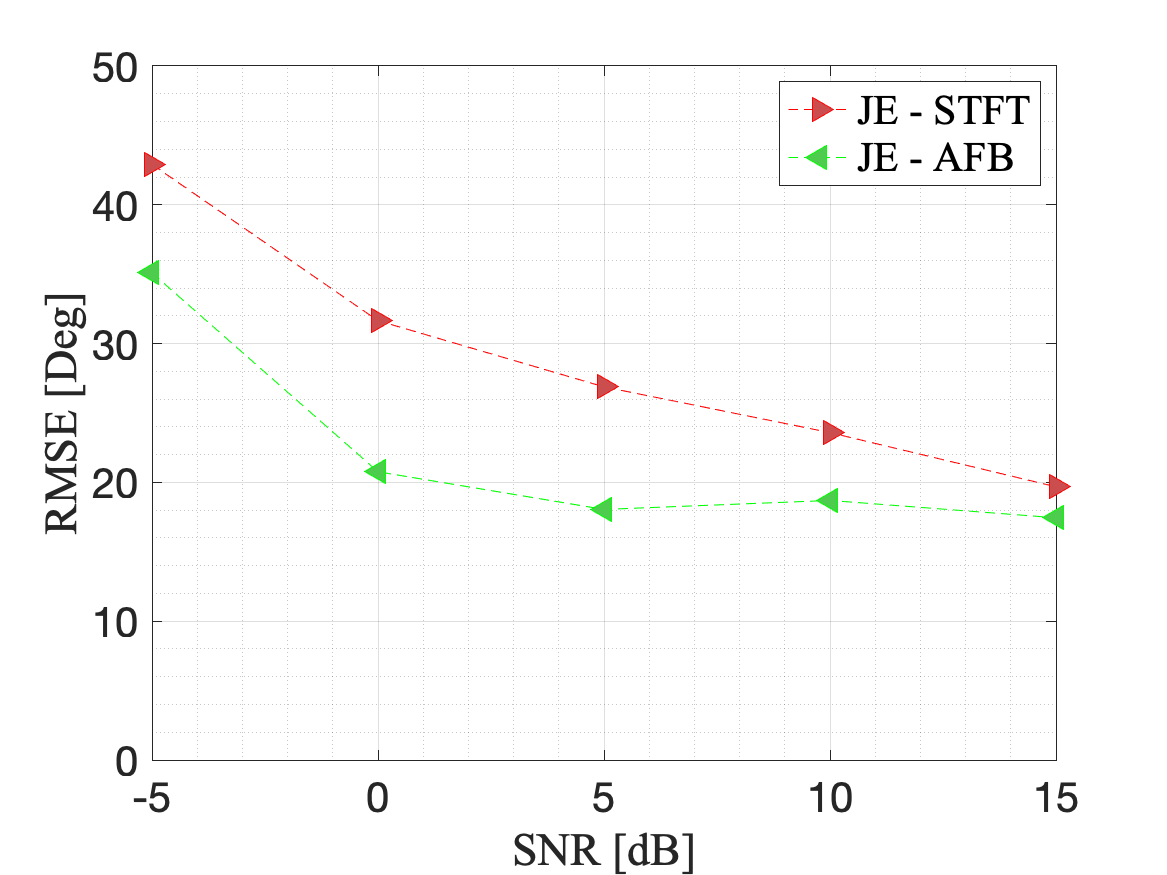}
         \caption{}
     \end{subfigure}
     \hfill
     \begin{subfigure}[b]{0.33\textwidth}
        \centering
        \includegraphics[width=\textwidth]{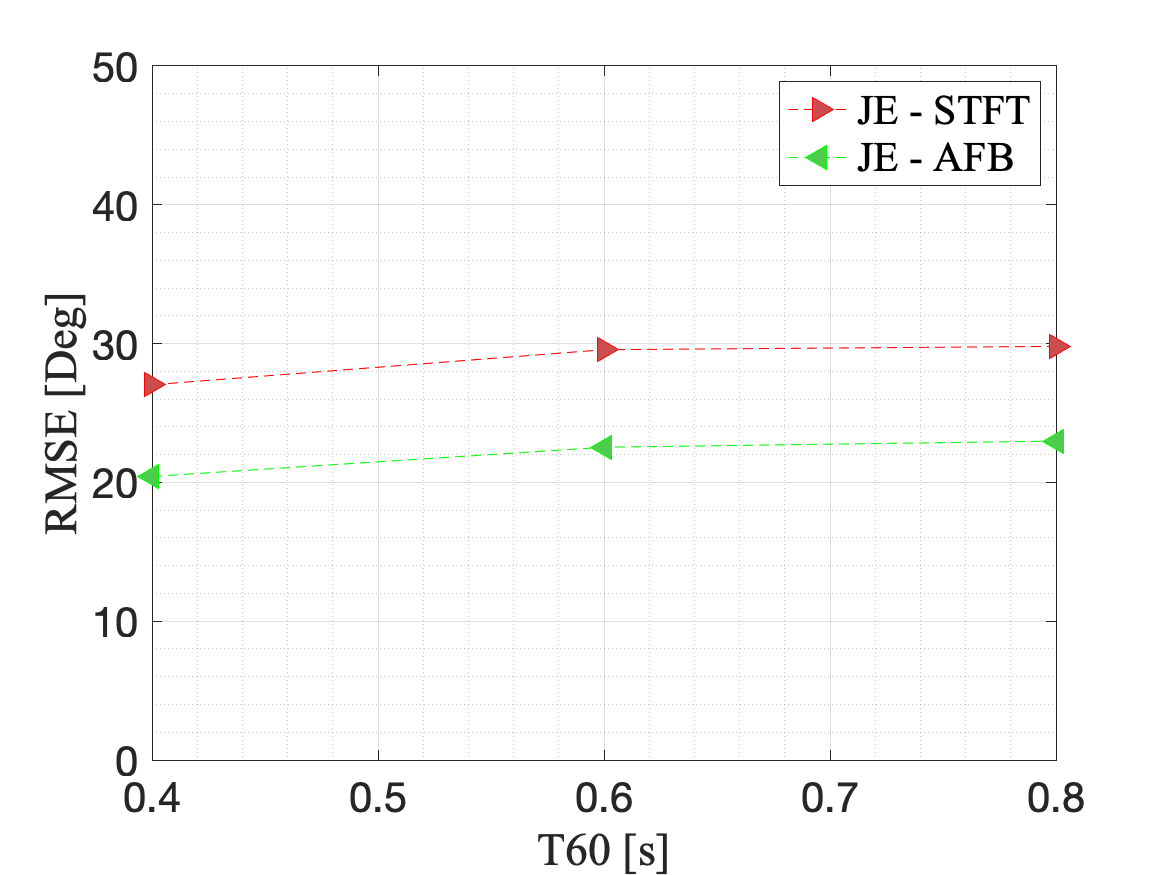}
        \caption{}
     \end{subfigure}
     \hfill
     \begin{subfigure}[b]{0.33\textwidth}
         \centering
         \includegraphics[width=\textwidth]{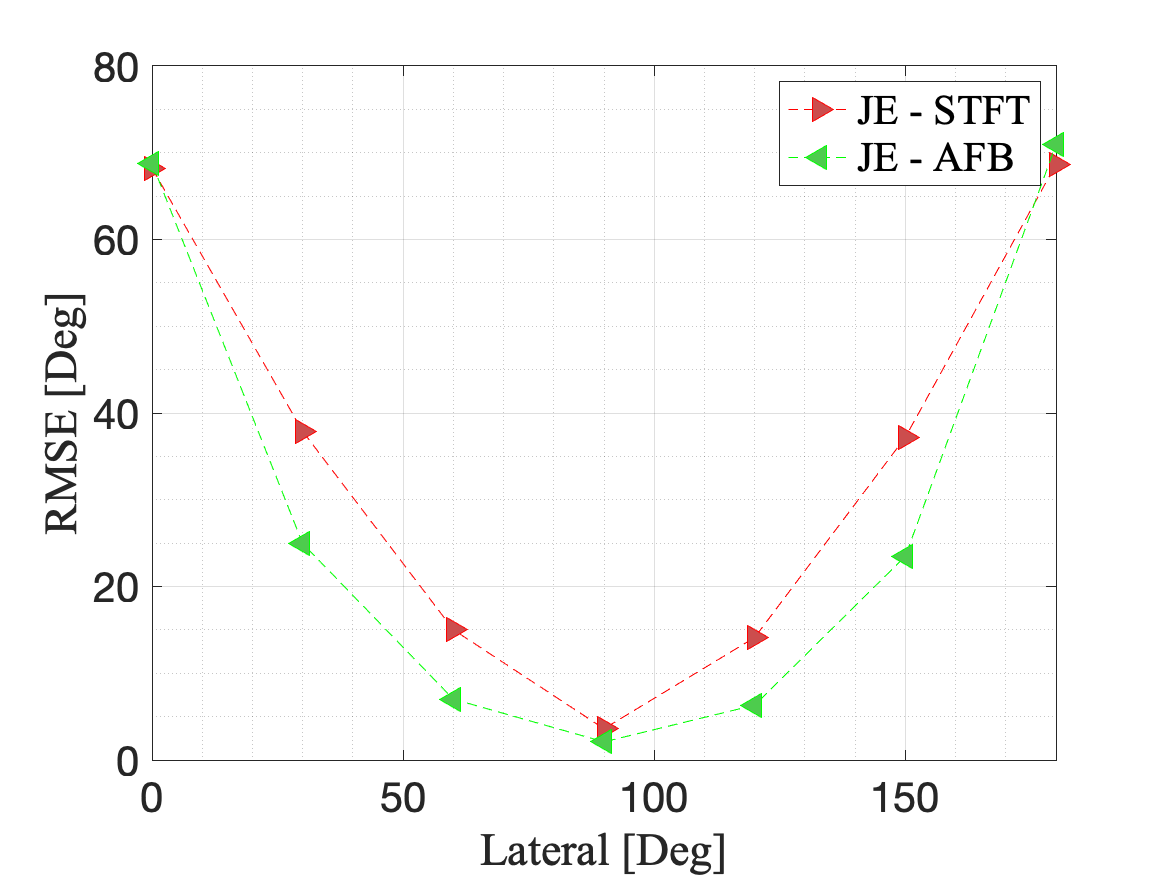}
         \caption{}
         \label{fig:DPD_JE_STFT2D_AUDFB2D_f}
     \end{subfigure}
     \hfill
    \caption {RMSE of the lateral estimation with the DPD and JE methods, using 2D angle search. Figure (a) shows the DPD method for different SNR (with randomly taken lateral angles and an average $T_{60}$ of $0.6$\,s). Figure (b) shows the DPD method for different $T_{60}$ (with randomly taken lateral angles and an average SNR of 5 dB). Figure (c) shows the DPD method for different lateral angles (with an average SNR of 5 dB and an average $T_{60}$ of $0.6$\,s). Figures (d-f) follow the same format as (a-c) respectively, but use the JE method instead.}
    \label{fig:DPD_JE_STFT2D_AUDFB2D}
\end{figure*}

DOA estimation performance is compared for the two methods, and when incorporating STFT versus auditory filters, and a 2D search versus a 1D lateral search.

For both methods, in the first stage of processing both STFT and auditory filters are applied. The STFT is computed with an FFT size of $1536$ samples ($32\,$ms) and a Hanning window with $50\%$ overlap. The auditory filters were computed with 42 frequency channels logarithmically spaced according to the ERB scale, between $60$ to $6000$ Hz, and sampled in time according to Eq. \eqref{eq:tauerb}. Performance of the methods based on the auditory filter bank will be referenced as AFB.

The DPD-test method is computed according to Section \ref{sec:DPDT}. The spatial correlation matrix is estimated according to Eq. \ref{eq:SpectrumMatrix} by averaging bins over time and over frequency. The number of bins in time is such that the averaging interval is equal to $64\,$ms, and the number of bins in the frequency is equal to 2. The threshold in Eq. \ref{eq:DPDT} is determined such that 5\% of the bins pass the test. In addition, TF bins from frequencies below 1kHz and beyond $6\,$kHz were excluded due to poor performance in those frequency regions \cite{BDPD}.
For every TF bins that passed the test, the MUSIC spectrum is calculated as in Eq.\ref{eq:MUSIC}, with the spectrum peak providing the DOA estimate.

According to the second DOA estimation method (the JE method), the ITD and the ILD are first estimated from the STFT or from the AFB. The time averaging scheme is the same as in the DPD method. The DOA is estimated by comparing the binaural cues from the STFT or AFB with the corresponding cue computed from the HRTF as a reference. There is no methodology for selecting good TF bins in the JE method because it was not originally designed to be reverberation robust. Therefore, the same bins from the DPD method were used here for the lateral angle estimation.

DOA estimation for both methods is computed in two ways. First, a complete 2D search is performed to estimate both azimuth and elevation, from which the lateral angle is computed \cite{INCSYS}. Second, as proposed in this paper, a 1D search is performed directly for the  lateral angle, with the derived lateral steering vectors, as in Eq. \ref{eq:LatSV}.

Finally, the lateral estimation is computed as the mean over all TF bins that passed the test, according to Eq. \ref{eq:MeanAz}. The root mean squared error (RMSE) is then calculated for each condition of the simulation. 

% \begin{figure}[t]
% \centering
% \includegraphics[width=\columnwidth]{figures/STFT.png}
% \caption{STFT magnitude of clean speech}
% \label{fig:STFT}
% \end{figure}

% \begin{figure}[t]
% \centering
% \includegraphics[width=\columnwidth]{figures/AUD_FB.png}
% \caption{AFB magnitude of clean speech}
% \label{fig:AUD_FB}
% \end{figure}

\subsection{STFT vs AFB}
\label{sec:Simu1}

\begin{figure*}[h!]
     \centering
     \begin{subfigure}[b]{0.33\textwidth}
         \centering
         \includegraphics[width=\textwidth]{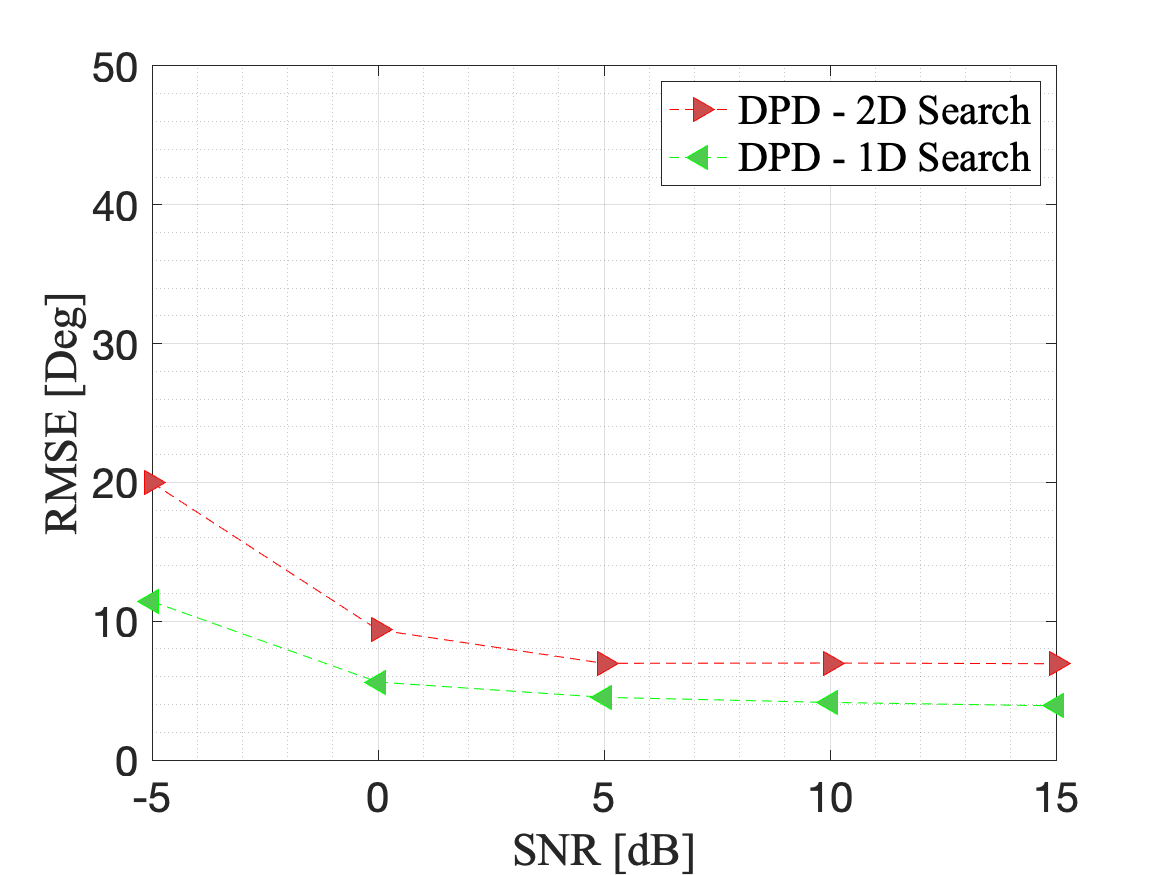}
         \caption{}
     \end{subfigure}
     \hfill
     \begin{subfigure}[b]{0.33\textwidth}
        \centering
        \includegraphics[width=\textwidth]{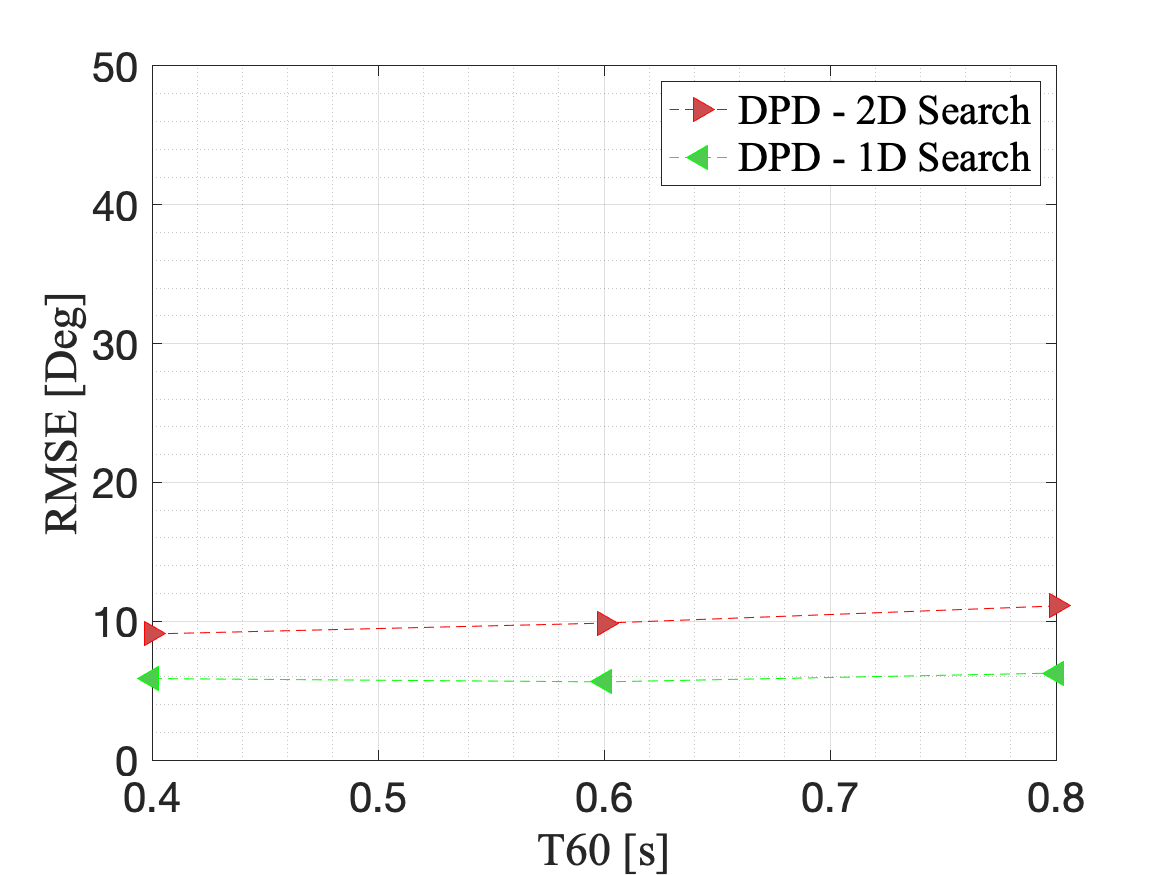}
        \caption{}
     \end{subfigure}
     \hfill
     \begin{subfigure}[b]{0.33\textwidth}
         \centering
         \includegraphics[width=\textwidth]{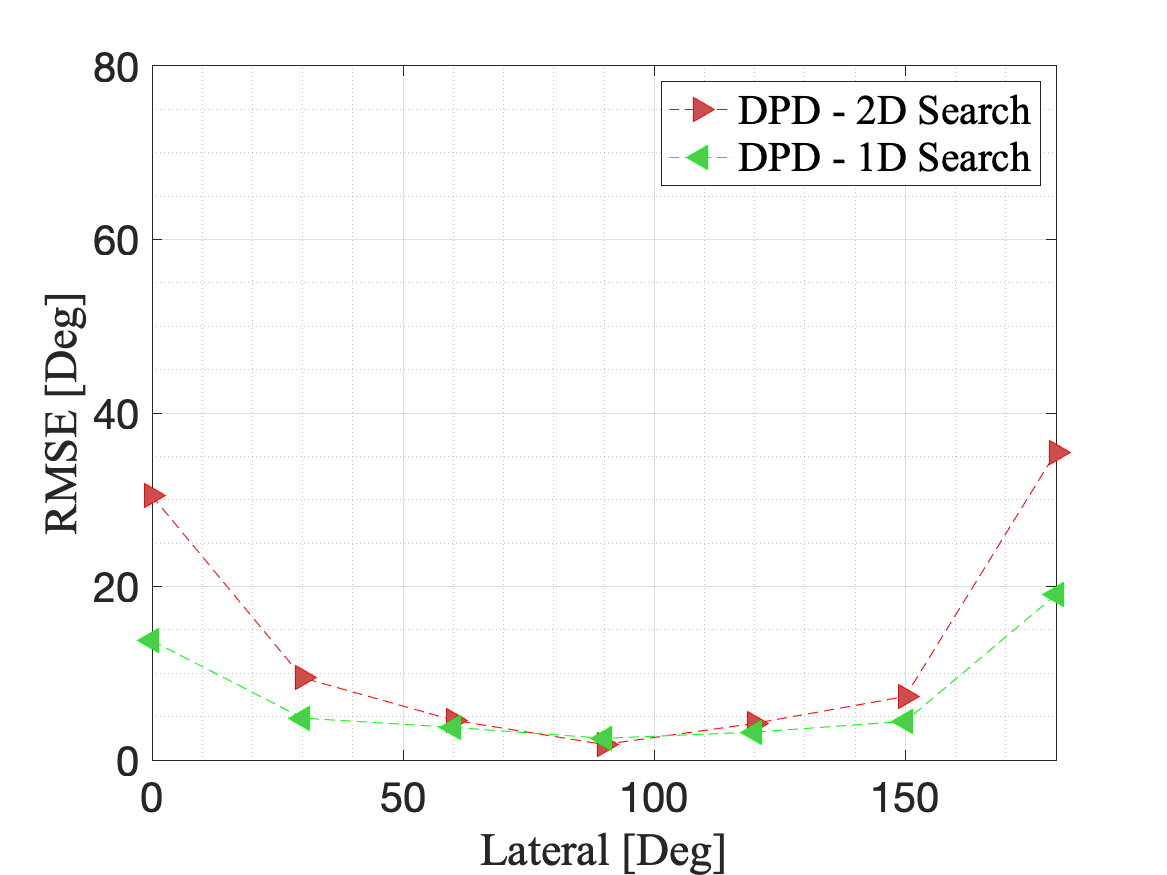}
         \caption{}
         \label{fig:DPD_JE_AFB2D_AFB1D_c}
     \end{subfigure}
     \vfill
     \begin{subfigure}[b]{0.33\textwidth}
         \centering
         \includegraphics[width=\textwidth]{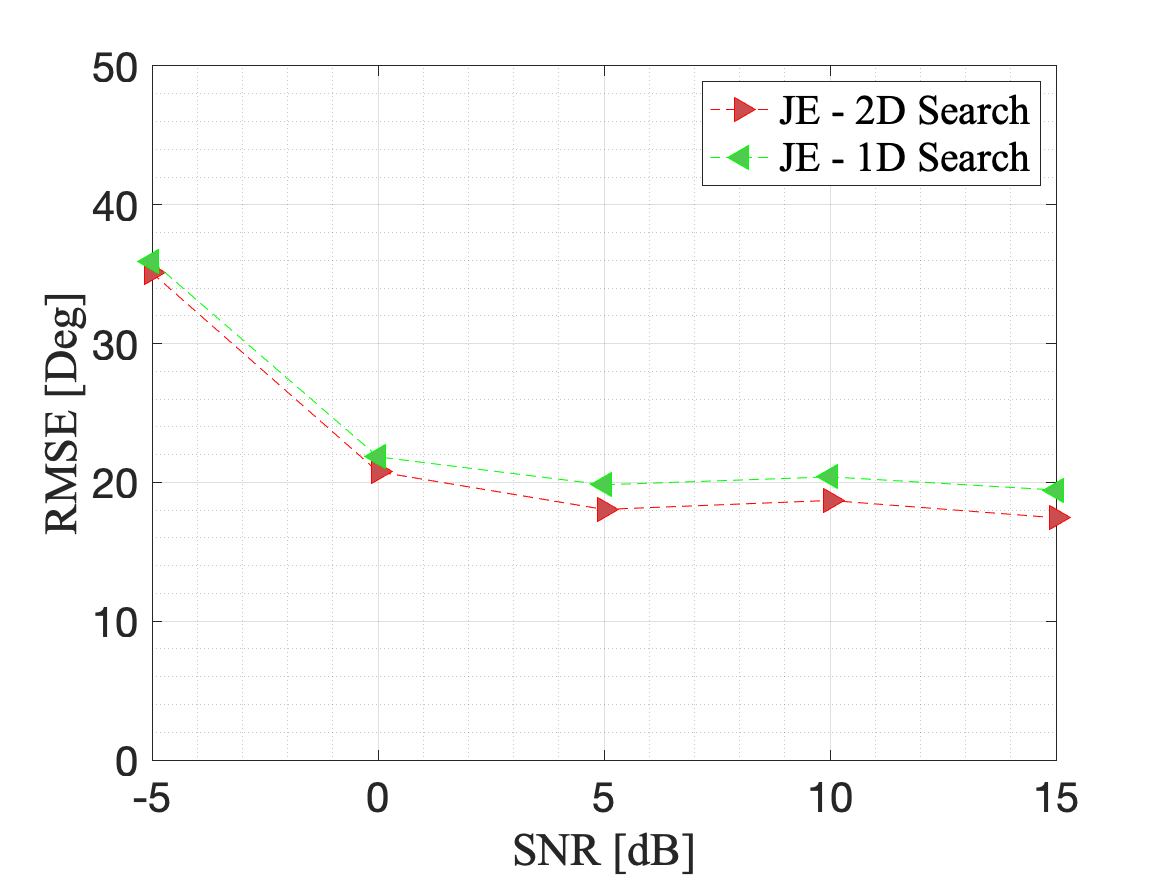}
         \caption{}
     \end{subfigure}
     \hfill
     \begin{subfigure}[b]{0.33\textwidth}
        \centering
        \includegraphics[width=\textwidth]{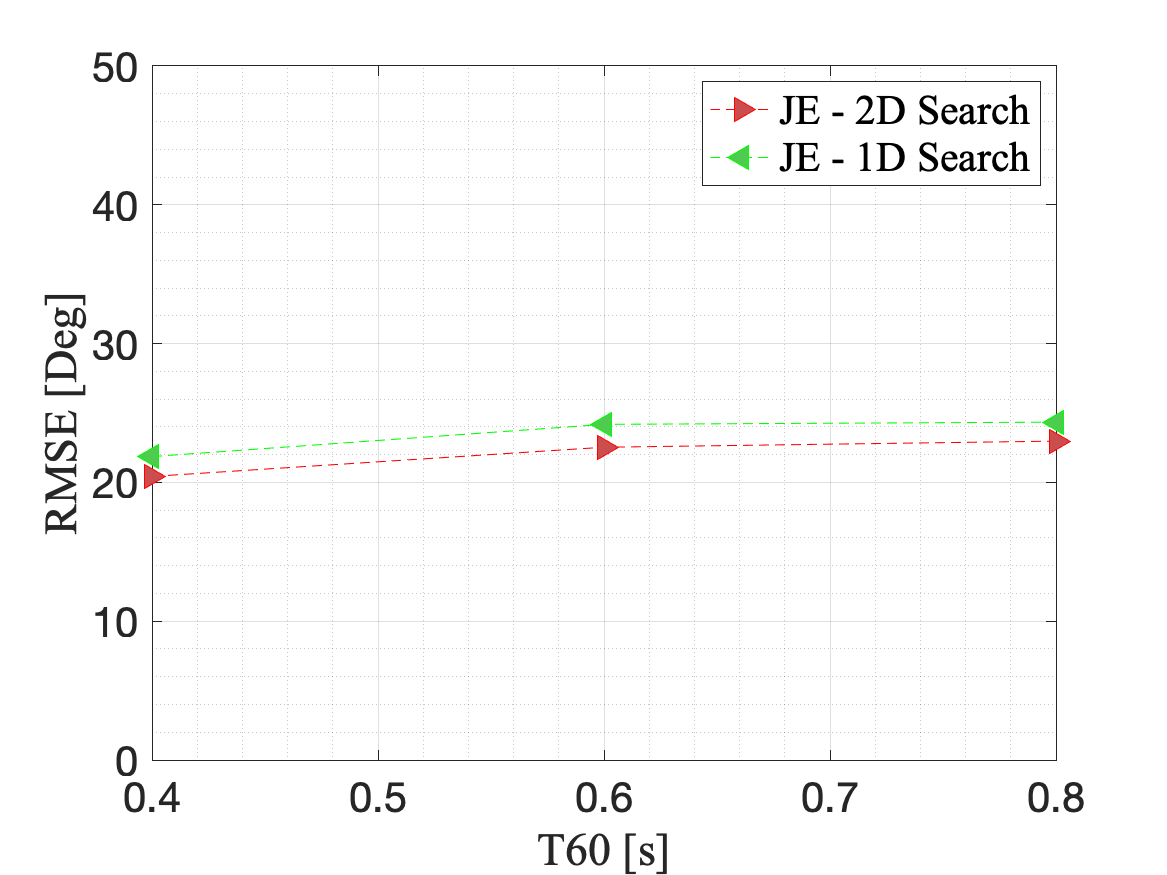}
        \caption{}
     \end{subfigure}
     \hfill
     \begin{subfigure}[b]{0.33\textwidth}
         \centering
         \includegraphics[width=\textwidth]{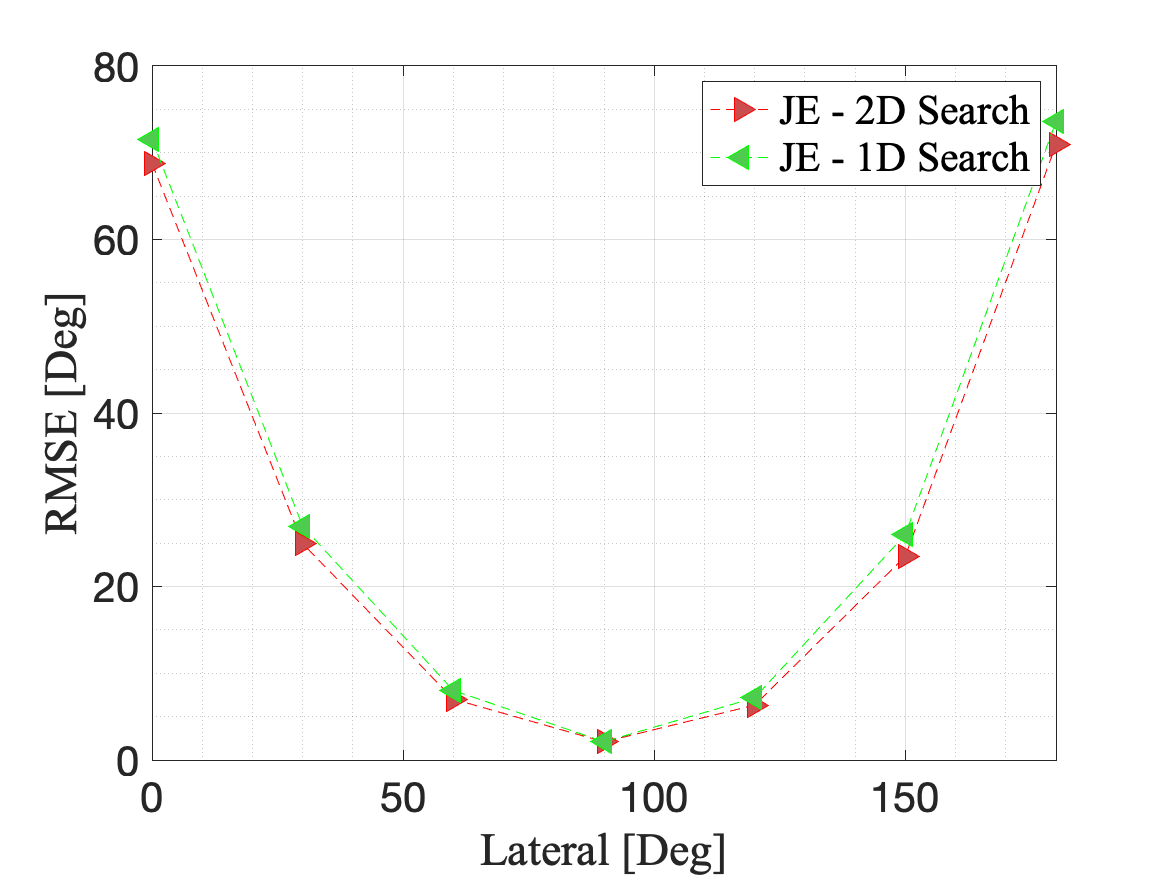}
         \caption{}
         \label{fig:DPD_JE_AFB2D_AFB1D_f}
     \end{subfigure}
    \caption {RMSE of the lateral estimation with the DPD and JE methods, using the AFB. Figure (a) shows the DPD method for different SNR (with randomly taken lateral angles and an average $T_{60}$ of $0.6$\,s). Figure (b) shows the DPD method for different $T_{60}$ (with randomly taken lateral angles and an average SNR of 5 dB). Figure (c) shows the DPD method for different lateral angles (with an average SNR of 5 dB and an average $T_{60}$ of $0.6$\,s). Figures (d-f) follow the same format as (a-c) respectively, but use the JE method instead.}
     \label{fig:DPD_JE_AUDFB1D_AUDFB2D}
\end{figure*}

In the first study, the performance of the methods when using STFT and AFB is compared. Figure \ref{fig:STFT} presents examples of STFT and AFB magnitude for clean speech, showing the spread of energy over both frequency and time in both representations. 

Fig. \ref{fig:DPD_JE_STFT2D_AUDFB2D} presents the RMSE of the lateral angle estimation for both the DPD and JE methods, averaged over all conditions. 
Figures (a), (d) represent the results for varying SNR; (b), (e) for varying $T_{60}$; and (c), (f) for varying lateral angle direction, for both STFT and AFB. In both cases a 2D search was used.
As expected, the figure shows that the error increases with longer reverberation time, lower SNR, and lateral directions away from the front ($90^{\circ}$) \cite{EdgesLat}.
The AFB seems to perform better than the STFT - the RMSE for the AFB is significantly lower then for the STFT, especially under worse conditions, i.e., low SNR, high $T_{60}$ and lateral directions away from the front.
In Fig. \ref{fig:DPD_JE_STFT2D_AUDFB2D}, the upper row (a-c) presents results for the DPD method and the lower row (d-f) presents results for the JE method. Both methods exhibit similar trends. 
In summary, frequency analysis with AFB seems to outperform the STFT for both methods, motivating the use of AFB for DOA estimation.

\subsection{2D vs 1D angle search}\label{sec:Simu 2}

\begin{figure}[ht]
\centering
\includegraphics[width=\columnwidth]{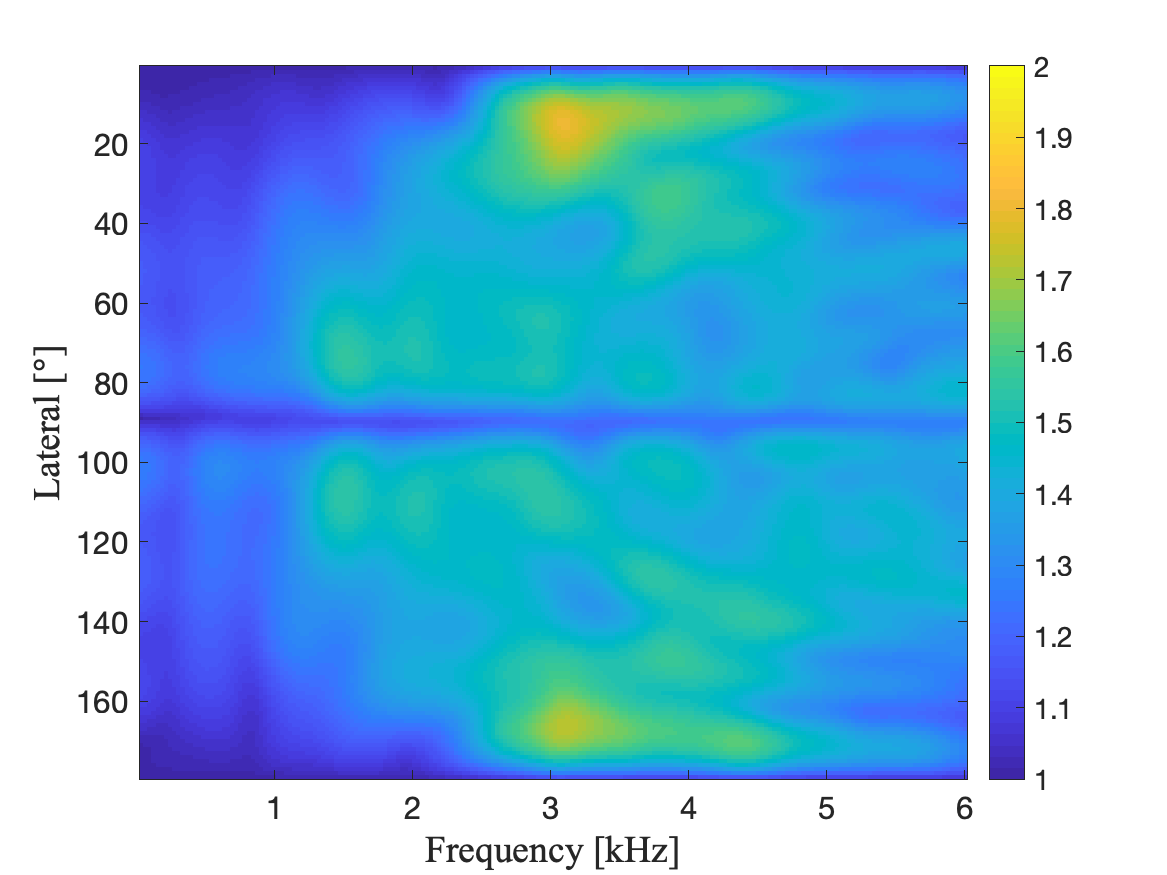}
\caption{Effective rank of steering matrix $\mathbf{H}(\theta^{'},\omega)$, defined in Eq. \ref{eq:HRTFMatrix} }
\label{fig:erank}
\end{figure}

In this study, the performance of the lateral angle estimation with 2D and 1D searches is compared. For this purpose, the lateral steering vectors are computed as in Eq. \ref{eq:LatSV}.
Figure \ref{fig:erank} presents the effective rank \cite{effective_rank} of the steering matrix $\mathbf{H}(\theta^{'},\omega)$ defined in Eq. \eqref{eq:HRTFMatrix} in the lateral angle - frequency domain. The figure shows that the effective rank is close to 1 in most lateral angle - frequency regions, due to the HRTF similarity within a cone. This supports the formulation of lateral angle steering vectors as in Eq. \ref{eq:HRTFMatrix}, which are based on a rank-1 approximation. 

Fig. \ref{fig:DPD_JE_AUDFB1D_AUDFB2D} presents the RMSE of the lateral angle estimation in a way that is similar to Fig. \ref{fig:DPD_JE_STFT2D_AUDFB2D}, but for 2D and 1D angle searches. The AFB is used in this case as it outperformed the STFT, as presented in the previous section.
In addition to requiring a more computationally efficient search due to the dimension reduction in the search grid,
the figure shows that the 1D search outperforms the 2D search with respect to RMSE for the DPD method.

For the JE method, similar results are obtained for the two search methods, while the 1D search is still preferred in terms of computation complexity. The latter can be explained by the use of ITD and ILD in this method, which inherently maps the lateral angle. In summary, the 1D search outperformed the 2D search for the DPD method, which is based on steering vectors, while for the JE method the 1D search only incorporated a more efficient search.

\subsection{Computational complexity}

This section aims to study the computational efficiency of the two search methods. The total running times of the entire algorithm, as detailed under Algorithm \ref{alg:doa_estimation}, is used as a measure for computation complexity. 
The methods were implemented in MATLAB (2022 version), running on a MacBook with 16 GB RAM and a 2.2 GHz Intel Core i7 processor. The average running times for a single realization were measured to be 867 ms for the 2D search and just 37 ms for the 1D search method, for a 5-second speech segment. This underlines the significantly lower computational demand of the 1D search.

\section{Experimental study with BRIR data}\label{sec:BrirStudy}
% For the STFT vs AFB chapter:
\begin{figure*}[h!]
      \centering
      \begin{subfigure}[b]{0.45\textwidth}
          \centering
          \includegraphics[width=\textwidth]{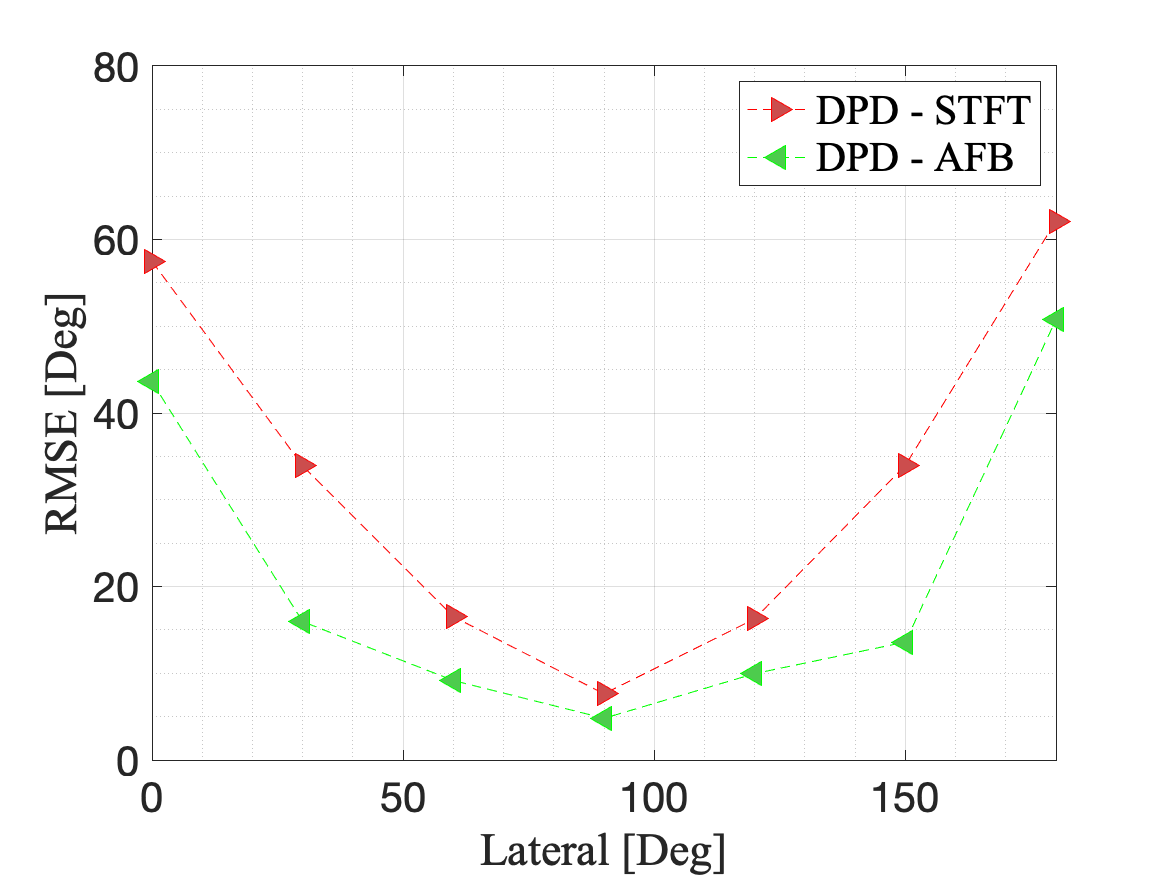}
          \caption{}
      \end{subfigure}
      \hfill
      \begin{subfigure}[b]{0.45\textwidth}
         \centering
         \includegraphics[width=\textwidth]{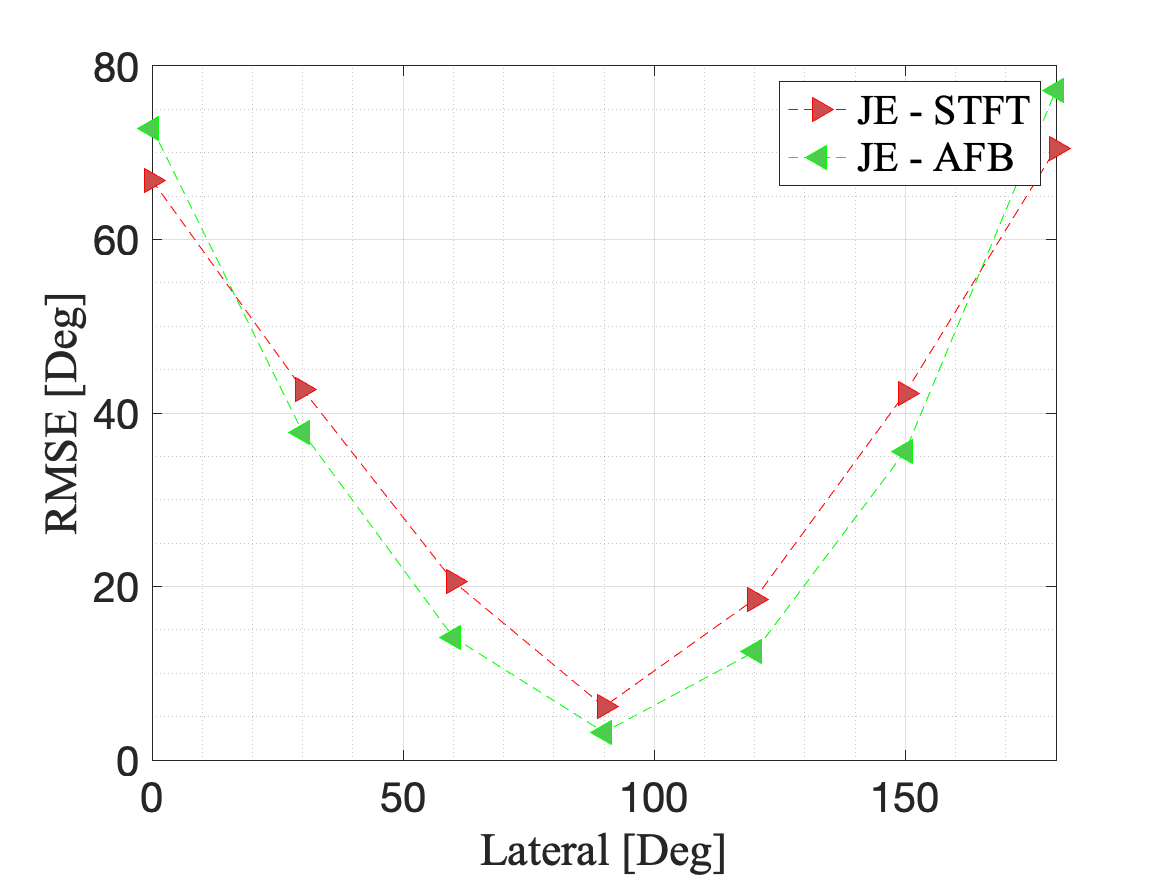}
         \caption{}
      \end{subfigure}
     \caption{RMSE of the lateral estimation with the DPD and JE methods, using 2D angle search. The figures show the results for different lateral angles with a $T_{60}$ of $0.27$\,s and SNR of 5 dB. Figure (a) presents results for the DPD method. Figure (b) presents results for the JE method.}
      \label{fig:BRIR_DPD_JE_STFT2D_AFB2D}
\end{figure*}

% For the 2D vs 1D Angle Search chapter:
\begin{figure*}[h!]
      \centering
      \begin{subfigure}[b]{0.45\textwidth}
          \centering
          \includegraphics[width=\textwidth]{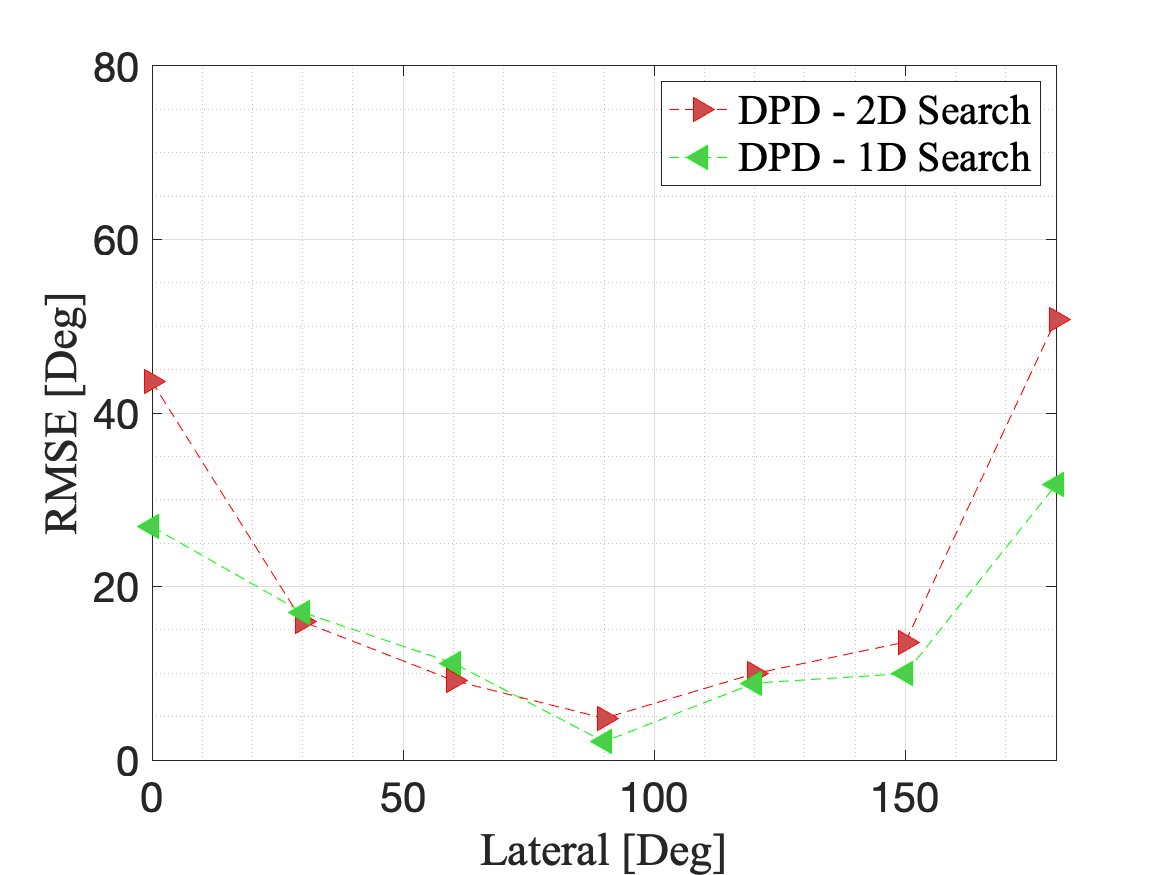}
          \caption{}
      \end{subfigure}
      \hfill
      \begin{subfigure}[b]{0.45\textwidth}
         \centering
         \includegraphics[width=\textwidth]{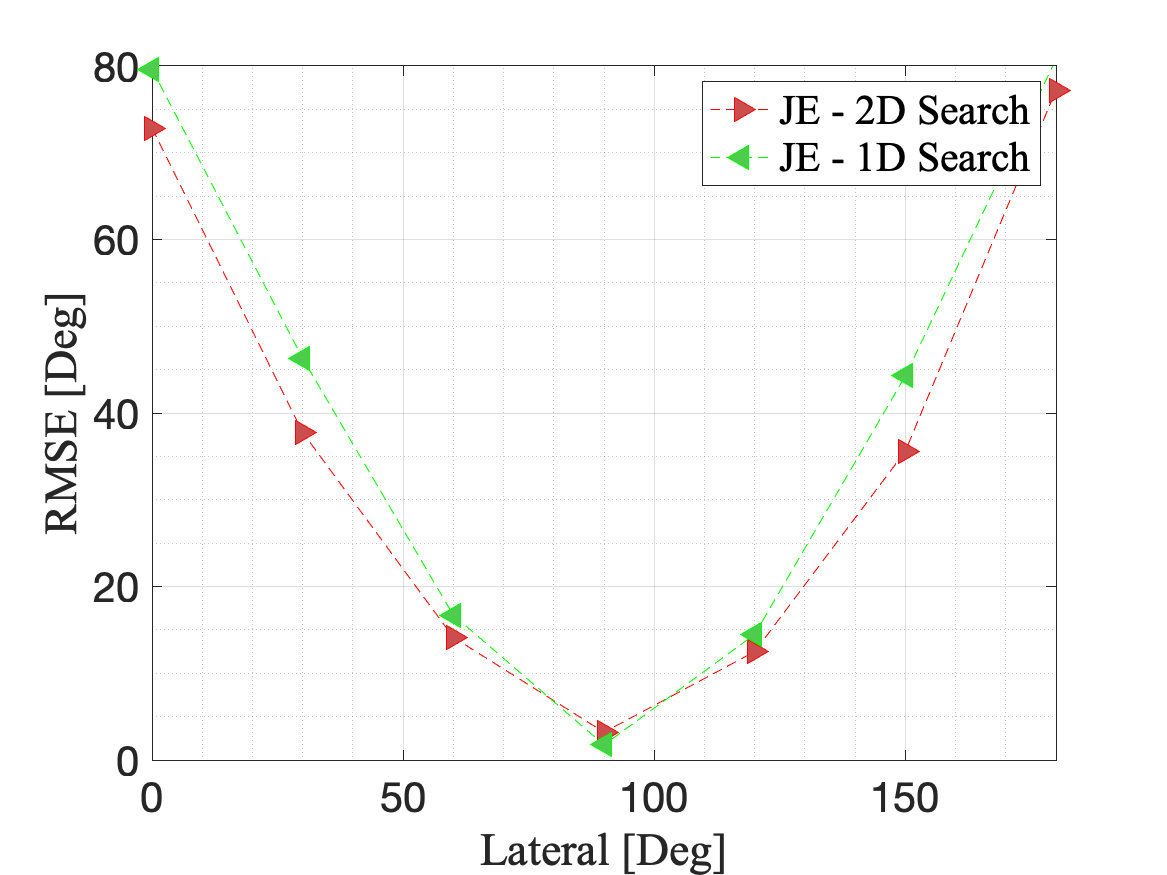}
         \caption{}
      \end{subfigure}
     \caption{RMSE of the lateral estimation with the DPD and JE methods, using the AFB. The figures show the results for different lateral angles with a $T_{60}$ of $0.27$\,s and SNR of 5 dB. Figure (a) presents results for the DPD method. Figure (b) presents results for the JE method.}
      \label{fig:BRIR_DPD_JE_AFB2D_AFB1D}
\end{figure*}

This chapter presents an experimental study  based on measured binaural room impulse responses (BRIR), aiming to validate the theory and simulation results. 

\subsection{Setup}
The experiment is based on a dataset derived from a library of BRIR, captured in a controlled environment at the University of Salford \cite{melchior2014library}. The experiment was conducted in a room of dimensions 5.8 $\times$ 6.6 $\times$ 2.8 m, with average reverberation time of 0.27s, and under signal-to-noise ratio of 90 dB. The recordings were performed using a sample rate of 48kHz.

In the dataset, BRIRs were measured with the sound source (loudspeaker) positioned at various directions  relative to the KEMAR manikin that was used to capture the binaural signals. A directional resolution of $2^{\circ}$ was used along the azimuth. The sound source was positioned in the horizontal plane, leading to elevation of $0^{\circ}$ between the source and the manikin. Among the 15 distinct manikin position, the central room position was chosen for this study. In this specific configuration, the distance between the source and the manikin was consistently 2.1 m.

To recreate audio signals that would have been captured in the room, audio files from the TIMIT database \cite{TIMIT} were used. For each realization, we randomly selected a speaker from a group of 2 male and 2 female speakers, consistent with the approach in the simulations setup (chapter \ref{sec:Setup}). The chosen audio signals were then convolved with the BRIRs to compute binaural signals.
Then, Gaussian white noise was added to the binaural signals, with a signal-to-noise ratio of 5 dB, in a way similar to the simulation study, to produce more realistic noisy signals.

The following results represent an aggregation of error analyses across different realizations taken from varying source locations and azimuth directions within the room, in a way that is similar to with the methodology described in the simulation study (chapter \ref{sec:Methodology}).

\subsection{STFT vs AFB}

In this section, performance of the algorithm with STFT and AFB is compared using the experimental data. Note that the SNR and reverberation time are fixed in this case. 
Fig. \ref{fig:BRIR_DPD_JE_STFT2D_AFB2D} presents the RMSE of the lateral angle estimation for both the DPD and JE methods.
The results show that the error with AFB is lower, especially for lateral directions away from the front ($90^{\circ}$) \cite{EdgesLat}, and especially for the DPD algorithm. The similarity between the simulated (Fig. \ref{fig:DPD_JE_STFT2D_AUDFB2D_c} and \ref{fig:DPD_JE_STFT2D_AUDFB2D_f}) and experimental (Fig. \ref{fig:BRIR_DPD_JE_STFT2D_AFB2D}) results, further validate the effectiveness of the AFB approach in DOA estimation.

\subsection{2D vs 1D angle search}

Fig. \ref{fig:BRIR_DPD_JE_AFB2D_AFB1D} illustrates
the RMSE of the lateral angle estimation for both 2D and 1D angle searches with the experimental data, for the DPD and JE methods.
The results shows that the trends align with the simulation findings as in Figs. \ref{fig:DPD_JE_AFB2D_AFB1D_c} and \ref{fig:DPD_JE_AFB2D_AFB1D_f}. The 1D search shows better results with the DPD method, in particular for the extreme lateral directions. However, for the JE method, the difference between the two searches seems less significant, with a slight advantage for the 2D search. Overall, this is consistent with the simulation results. With the 1D search providing similar or better performance compared to the 2D search, its reduced computational saving becomes an advantage.

\section{Conclusions}\label{sec:Conclusions}

This paper proposed and investigated new alternatives for processing in binaural DOA estimation, which included the incorporation of an auditory filter bank, and direct lateral angle estimation. These processing alternatives have been theoretically developed and incorporated into the DPD and the JE methods as examples. The proposed alternatives outperformed the original methods in most cases. The study suggests that improved performance is achieved with the proposed alternatives in terms of DOA estimation error and computational efficiency, and can also be generalized to other binaural localization methods.

\section*{Acknowledgments}\label{sec:Acknowledgments}

This research was supported by THE ISRAEL SCIENCE FOUNDATION under Grant 966/18.

\bibliographystyle{elsarticle-num} 
\bibliography{mainref}
\end{document}